\RequirePackage{fix-cm}

\documentclass[smallextended]{svjour3}       
\smartqed  
\usepackage{graphicx}
\usepackage{natbib} 
 \usepackage{mathptmx}      
\usepackage{latexsym}
\usepackage{amssymb}
 \journalname{Rheologica Acta}
\begin{document}

\title{ Fractures in Complex Fluids: the Case of Transient Networks}


\author{Christian LIGOURE         \and
        Serge MORA 
}


\institute{C. Ligoure \at
              Laboratoire Charles Coulomb - UMR 5221\\
              Universit\'e Montpellier 2 and CNRS\\
              Place E. Bataillon. F-34095 Montpellier Cedex, France
              \email{christian.ligoure@univ-montp2.fr}           
           \and
           S. Mora \at
           Laboratoire Charles Coulomb - UMR 5221\\
           Universit\'e Montpellier 2 and CNRS\\
           Place E. Bataillon. F-34095 Montpellier Cedex, France
           \email{serge.mora@univ-montp2.fr}
}

\date{Received: date / Accepted: date}

\maketitle

\begin{abstract}
We present a comprehensive review of the current state of fracture phenomena in transient networks, a wide class of viscoelastic fluids. We will first define what is  a fracture  in a complex fluid, and recall the main structural and rheological properties of transient networks. Secondly, we review experimental reports on fractures of transient networks in several configurations: shear-induced fractures, fractures in Hele-Shaw cells and fracture in extensional geometries (filament stretching rheometry and pendant drop experiments), including fracture propagation. The tentative extension of the concepts of brittleness and ductility to the fracture mechanisms in transient networks is also discussed. Finally, the different and apparently contradictory theoretical approaches developed to interpret fracture nucleation will be addressed and confronted to experimental results. Rationalized criteria to discriminate the relevance of these different models will be proposed.


\end{abstract}

\newpage
List of symbols\\

\begin{tabular}{ll}
$G_0$ &shear modulus\\
$\tau$ &relaxation time\\
$\eta$ & viscosity\\
$\gamma$& shear strain\\
$\dot \gamma$& shear rate\\
$\dot \epsilon$& elongational strain rate\\
$\sigma$& stress\\
$\sigma_{xy}$ & shear stress \\
$N_1$ or $\sigma_N$& First normal stresses difference\\
$W$& Griffith energy cost\\
$L$&crack size\\
$L_c$ & Griffith length\\
$F_s$& interfacial cohesive energy per unit area\\
$t_b$& delay (waiting) fracture time\\
$\langle t_1\rangle $&average delay  (waiting) fracture time  predicted by the {\em Thermally activated crack nucleation model}\\
$\langle t_2 \rangle$&average delay (waiting) fracture time predicted by the {\em Activated bond rupture model}\\
$\langle t_3\rangle $&average delay (waiting) fracture time predicted by the {\em Self healing and activated bond rupture nucleation model}\\
$\sigma_c$& failure stress\\
$\sigma_1$& characteristic stress involved in the {\em Thermally activated crack nucleation model}\\
$\sigma_2$&characteristic stress involved in the {\em Activated bond rupture model}\\
$\sigma_3$&characteristic stress involved in the {\em Self healing and activated bond rupture nucleation model}\\
${\cal G}$& strain energy release\\
$V$&tip velocity\\
$k_B$& Boltzmann constant\\
$T$& temperature\\
$\xi$& typical distance between junctions in a network\\
$\rho$&density\\
$Q$&injection rate\\
$\delta$&length of a solvophobic end\\
$De$&Deborah number\\
$\Gamma $&Interfacial  or surface tension \\
\end{tabular}

\section{Introduction}
\label{intro}

The ability of viscoelastic fluids to fracture has been recognized in the sixties [\cite{Hutton1963}] but remains much less documented that the breakdown of solid materials. The starting point for discussing fracture in viscoelastic fluids is coming up with a rigorous definition of fracture. The layman's definition of failure  proposed for solid materials by Buehler [\cite{Buehler2010}] \-- {\em failure occurs when the load bearing capacity of the material under consideration is significantly reduced or completely lost due to a sudden development} \-- is not applicable for fluids, which  accommodate arbitrarily large deformations after a finite time. At a more fundamental level Buehler argues that fracture of a material due to mechanical deformation can be understood as conversion of elastic energy  into breaking of chemical bonds or heat. This definition is more appropriate for fluids but does not allow to discriminate clearly the fragmentation of liquids due to some hydrodynamic  instability  [\cite{Eggers2008}] and the fracture of fluids reminiscent of the fracture of solids. For instance, liquid jets serve as a paradigm for hydrodynamic instability leading to drop breakup through the Rayleigh-Plateau capillary instability: in this case, intermolecular (elastic) energy is converted at the surface vicinity alone. A tentative definition for the fracture of fluid  could be the following:" the fracture of  a  fluid due to mechanical deformation, including flow, can be understood as a dissipation of \textit{bulk} elastic energy into  breaking of physical or chemical bonds." This definition excludes  the hydrodynamic instabilities observed also in inviscid or Newtonian fluids of the  field of fluid's fracture because elastic energy that come in play is of interfacial  nature solely. 

Complex fluids denotes a (too)  huge class of  condensed-phase materials that posses mechanical properties intermediate between ordinary liquids and ordinary solids [\cite{Larson1999}]. Many of them are "solids" at short time and "liquid "at long time, hence they are \textit{viscoelastic}; the characteristic time required of them to change from "solid" to "liquid" varies from fractions of a second to  hours. Numerous extensive studies on the rupture of  entangled polymer melts  in extension have been carried out in the past forty years. This allows to distinguish two zones of either viscoelastic rupture or elastic fracture in the Malkin-Petrie master curve  intended to illustrate the different extensional responses with increasing strain rates of entangled polymer melts [\cite{Malkin1997}].  These two types of rupture have been recently  reconsidered by Wang  [\cite{Wang2010}]: a yield-to-rupture failure transition  is observed: the yield  failure (called also ductile failure  by other authors [\cite{Ide1976,Ide1977,Ide1978}] should be due to the yielding of entanglements, whereas at higher extensional strain rates,  the  specimen  breaks up in a purely elastic regime (without any flow).

Among visoelastic fluids, self-assembled  transient networks, that consists of reversibly cross-linked polymers in solutions constitute model systems for the physicist, with well defined structural properties and simple linear rheological behavior. The aim of this review is to embrace   the current aspects of the phenomenology of transient network's fractures in a compressive manner. We first present  what are transient networks: systems and rheology. Second, we report on  fracture's experiments  and computer simulations of transient networks  in several configurations: shear geometry, Hele-Shaw cells and extensional geometry. This also includes, fracture propagation and the extension of the concepts of brittleness and ductility for viscoelastic fluids. Then, we describe the several  theoretical approaches  developed to describe crack nucleation in transient networks, and how they compare to experiments, before to conclude.

\section{Transient networks: systems and rheology}
\label{sec:1}
 Self-assembled transient (often called physical) networks   are a class of complex materials forming spontaneously 3D networks at thermodynamical equilibrium, that can  transiently transmit  elastic stresses over macroscopic distances. Transient self-assembled networks are common in both natural and synthetic materials. They consist of reversibly cross-linked polymers  in which weak interactions such as hydrogen bonds, hydrophobic interactions, van der Waals forces, or electrostatic interactions are responsible for cross-links formation. Because of the transient character of the junctions, and so of the thermodynamics equilibrium state of these systems at rest,  they don't  exhibit any  aging nor yield stress contrary to other classes of soft  out-of- thermodynamic equilibrium viscoelastic materials like dense particulate suspensions (colloidal glasses), or thermoreversible gels [\cite {Larson1999}].  One of the major issue of transient polymer networks is to convey useful rheological properties to solutions, such as increased viscosity, gelation, shear-thinning or shear-thickening.  They can be used as controlled drug delivery systems [\cite{Sutter2007}],  rheological regulators in polymer blends [\cite{Kim2004}],  coatings, food, and cosmetics, or as matrix materials for tissue engineering [\cite{Kim1998}]. Self-assembled transient networks  consist mostly of binary solutions of associative polymers, or ternary solutions of associative polymers and self-assembled surfactant aggregates  even if supramolecular reversible networks formed by mixtures of small molecules associating by directional interactions has been also reported [\cite{Cordier2008}]. Associating polymers are macromolecules with a part that is soluble in a selective solvent (often water), the so-called backbone or spacer to which two or more moieties that do not dissolve in this solvent, the stickers, are attached. The stickers may be randomly distributed along the backbone or may be grouped in blocks. the association of such polymers in solution has been studied extensively, and many reviews can be found in the literature, see for example  [\cite{Larson1999,Winnik1997,Berret2003,Meng2006,Chassenieux2011}]. Note that the restrictive  definition of transient networks that we propose, excludes simple  entangled  normal or living polymer solutions and melts, where entanglements of polymer chains play the role of transient cross-links and exhibit elastic yielding [\cite{Malkin1997,Boukani2009}]. Telechelic polymers are often used as model linkers because they are architecturally simple: they consist of a long solvophilic mid-block with each end terminated by a solvophobic short chain (a sticker)  [\cite{Semenov1995}]. The stickers incorporate into the solvophobic domains of the aggregates and can bridge them via their solvent-soluble mid-block resulting in an attractive interaction between the aggregates. . The nature and the morphologies of the aggregates forming the network's junctions are versatile: (i) telechelic polymers in binary solution  that self-assemble spontaneously into non-interacting flower-like micelles at low concentration and form three dimensional networks above a percolation concentration [\cite{Annable1993,Serero2000,Werten2009,Seitz2006}], (ii) surfactant vesicles [\cite{Lee2005}], (iii) lyotropic lamellar phases [\cite{Warriner1997}]  (iv) wormlike micelles [\cite{Ramos2007,Lodge2007,Nakaya2008,Tixier2010}], (v) spherical micelles [\cite{Appell1998,Tixier2010}], (vi) oil-in-water [\cite{Bagger1997,Filali1999}]  or (vii) water-in-oil microemulsion droplets  [\cite{Odenwald1995}], (viii)  photocrosslinkable nano-emulsions [\cite{Helgeson2012}].
  
  The rheology of transient networks is determined by the amount and the life time of the bridges. In the simplest case  when the bridging chains are flexible and untangled, each bridging chain contributes about 1 $k_BT$ per unit volume  to the elastic modulus $G_0$ according to the theory of rubber elasticity [\cite{Green1946,Tanaka1992,Yamamoto1956}]. The typical value of the shear modulus for these networks ranges between few Pa to  several ten thousands Pa. The elastic response relaxes when the solvophobic blocks escape from the core. Often the escape is characterized by a single relaxation time so that the terminal relaxation of the viscoelastic properties is characterized by a single Maxwell process well separated from the faster internal modes that characterize the conformational relaxation of the chains [\cite{Annable1993,Serero2000,Michel2000,Filali2001,Tabuteau2009,Hough2006}]. Far above the percolation concentration, the viscoelastic relaxation time $ \tau$ is related to the average lifetime of a connection [\cite{Green1946}] which in turn depends on the breakage probability of a cross-link. The average lifetime of bridge will depend on its chemical nature, the external conditions and the physical state  of the cross-links. It can vary through few ms [\cite{Tixier2010}],  few seconds  [\cite{Michel2000,Filali2001}], up to hours [\cite{Serero2000,Skrzeszewska2010,Seitz2006}]. 
The simplicity of the linear viscoelastic behavior of most of self-assembled   transient networks is in contrast with their highly complex non-linear response that can vary widely [\cite {Pellens2004}]. Among all systems, the experimental model transient network consisting of oil-in-water microemulsion droplets reversibly linked by telechelic polymers [\cite{Filali2001,Michel2000}] is perhaps the unique one which exhibits the  steady shear flow curve of a pure Maxwell fluid for both the shear stress and the first normal stress difference until it breaks [\cite {Tabuteau2009}]. In most other systems, the steady shear viscosity exhibits three flow regimes. Above a Newtonian plateau (linear regime), the viscosity increases considerably (shear thickening) prior to the onset of shear thinning at higher shear rates [\cite{Annable1993,Xu1996,Otsubo1999,Berret2001,Pellens2004b,Tripathi2006}]. However, the shear thickening region is not always present [\cite{Michel2000,Tixier2010,Tirtaatmajda1997,Mewis2001,Skrzeszewska2010}]. A large number of constitutive theoretical models  are based on the temporary-network kinetic model for telechelic polymers  network theory [\cite{Tanaka1992}] by applying the original ideas formulated by Green and Tobolsky [\cite{Green1946}] and Yamamoto [\cite{Yamamoto1956}]
   and have been developed to tentatively capture the main features of non-linear rheological properties of these networks [\cite{Marrucci1993,Ahn1995,VandenBrule1995,Vaccaro2000,Tripathi2006,HernandezCifre2003}].
   
 \section{Fracture experiments} \label{sec :experiments}
 Several geometries have been considered to explore fracture mechanisms in transient networks: elongational flows using either extensional rheometer [\cite{Tripathi2006}], pendant drop experiments [\cite{Tabuteau2009,Tabuteau2011}], shear flows in rheometric  cells [\cite{Berret2001,Tabuteau2009,Tixier2010,Skrzeszewska2010}] or flows  in Hele-Shaw cells [\cite{Zhao1993,Ignes-Mullol1995,Vlad1999,Mora2010}].
 
 \subsection{Shear-induced fractures} \label{sec : shear}
 \subsubsection{Stationary shear rate} \label{sec : stationary}

\begin{figure}
  \centering
  \includegraphics[width=12cm]{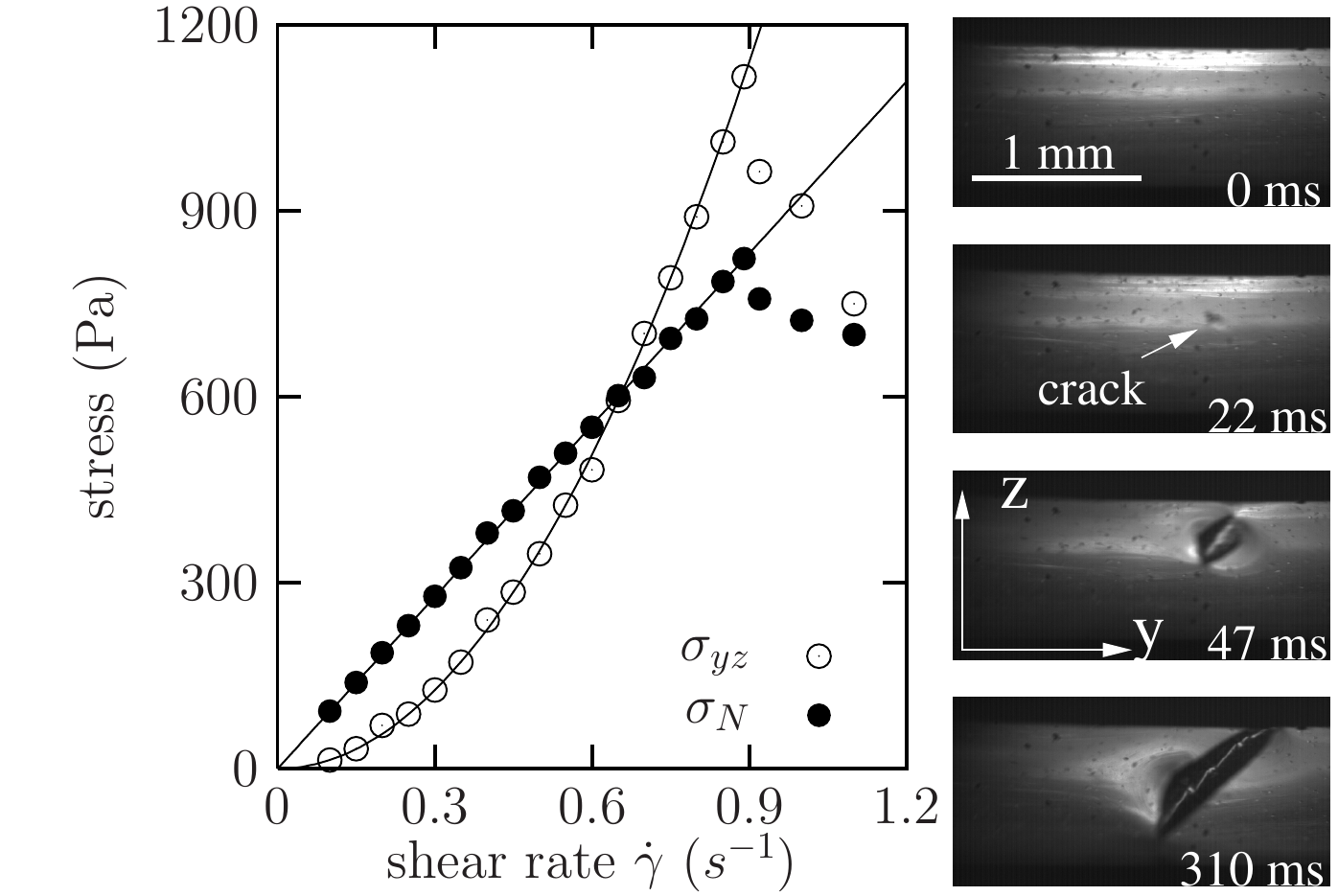}
\caption{ From [\cite{Tabuteau2009}]: \textbf{(Left)} shear stress ($\sigma_{yz}$) and first normal stresses difference  ($\sigma_N=\sigma_{zz}-\sigma_{yy}$) versus shear rate for a bridged microemulsion with a shear modulus $G_0=1210~Pa$ and a relaxation $\tau=0.8~\mathrm{s}$. The continuous lines are fits (Maxwell model). \textbf{(Right)} Development of a fracture, occurring at the surface of the sample, for a shear rate equal to $0.9 \mathrm{s}^{-1}$, corresponding to a critical  first normal stresses difference $\sigma_N=1190$ Pa.} 
\label{figure_flowcurve}       
\end{figure}

 When submitted to a constant shear rate  $\dot{\gamma}$, the measured shear stress  $\sigma_{xy}$ of "brittle" transient networks first increases smoothly with the shear rate [\cite {Molino2000,Berret2001,Tabuteau2009,Skrzeszewska2010,Tixier2010}]. However, above a critical shear rate $\dot{\gamma}\sim \tau^{-1}$, where $ \tau$ is the relaxation time of the network, the flow curve exhibits a sharp discontinuity (Figure \ref{figure_flowcurve}) and  the viscosity decreases abruptly. Below this value, the branch of the flow curve can be  Newtonian and shear thickening [\cite{Berret2001}], or  only Newtonian [\cite {Molino2000,Tabuteau2009,Skrzeszewska2010}], or Newtonian and shear thinning [\cite{Tixier2010}] depending on the   experimental system under consideration. Note that for "brittle "transient networks, the first normal stresses difference $N_1=\sigma_{xx}-\sigma_{yy}$  drops suddenly  for the same critical shear stress [\cite{Tabuteau2009}]. Authors of [\cite{Molino2000}] were the first to suggest that the sudden drop of the stress in the flow curve of  a transient network is the sign of a fracture propagation. 

 However, the first unambiguous demonstration of shear-induced fractures in transient networks  was done in [\cite{Berret2001}], by using a flow visualization technique in a  plate-plate transparent shearing cell.  At low shear rate, the velocity profile is homogeneous: the velocity decreases linearly from the rotating wall to the stationary one. Above the critical shear rate, the stationary velocity field within the gap of the cell exhibits a discontinuity, that defines a zone of fracture as the part of the fluid submitted to a high shear rate ($\sim10$ times the applied rate). This has been confirmed by Skrzeszewska {\em et al} [\cite{Skrzeszewska2010}] by Particle Image Velocimetry: the fracture zone has an irregular shape and is rather wide on the order of a few hundred micrometers. With increasing overall shear rate, the width of the fracture zone increases. The fracture zone can happen everywhere in the gap and is different for every experiment. For non-adhesive gels, the fracture zone occurs generally at one of the wall [\cite{Berret2001}], and so appears as sliding, that is observed in other classes of complex fluids like pastes or concentrated emulsions [\cite{Meeker2004}], whereas for adhesive gels [\cite{Skrzeszewska2010}], or rough wall surfaces [\cite{Tixier2010}], the fracture occurs in the bulk. A direct optical observation of the shear fractures [\cite{Tabuteau2009}] shows that above the critical stress, cracks open up all around the sample and grow rapidly. It is worth noting that the fractures are tilted  $45^{\circ}$ from the shear plane, perpendicular to the direction of the maximum extension (Figure \ref{figure_flowcurve}). Interestingly, for two very different systems [\cite{Tabuteau2009,Skrzeszewska2010}] that have perfectly Newtonian flow curve, before fracture occurs, the fracture shear stress in the steady state flow curve scales as $\sigma_{xy}\sim G_0$, where $G_0$ is the shear modulus of the network. In contrast to solids, here the fractures heal over rapidly after the shear rate is switched off, and a new experiment can be performed after a few minutes with quantitatively the same behavior [\cite{Tabuteau2009,Skrzeszewska2010}].
 
 Above the critical fracture shear rate, strong fluctuations of the shear stress have been observed [\cite{Sprakel2009b,Tixier2010,Ramos2011}], that can lead to an apparent shear plateau [\cite{Sprakel2009b}] reminiscent of shear-banded flows observed in a wide variety of soft materials such as solutions of  entangled wormlike micelles, colloidal suspensions and entangled polymer solutions [\cite{Fielding2007,Olmsted2008}]. Such strong fluctuations of the shear stress have also been observed  in solutions of entangled wormlike micelles at high shear rate and have been associated also with a rupture-like behavior as evidenced by Particle Tracking Velocimetry [\cite{Boukany2007}]. By using a novel class of transient networks made of surfactant micelles of tunable morphology [\cite{Tixier2010}] (from spheres to rods to flexible worms) linked by telechelic polymers, and coupling rheology and time-resolved structural measurements,  Ramos and Ligoure [\cite{Ramos2011}]  clearly show that true shear-banding is \textit{not} associated with strong fluctuations of the shear stress. Indeed, fluctuations  of the shear stress are entirely correlated to the fluctuations of the degree of  alignment of the micelles that can probe a fracture process. Sprakel {\em et al} [\cite{Sprakel2009b}] argue that the intermittent behavior observed in the stress response is due to repeated microfracture-healing events in the material. The cumulative distribution of the total stress drops $\Delta\sigma$ during a fracture displays a characteristic power-law behavior, ${\cal P}(>\Delta\sigma)\propto \delta\sigma^{-0.85}$, with ${\cal P}$ the stress probability distribution. The exponent is close to the value of 0.8 reported for true stick-slip motion [\cite{Feder1991}]. Attempts to explain such scaling behavior involve the concept of self-organized criticality.
   \begin{figure}
   \centering
  \includegraphics[width=12cm]{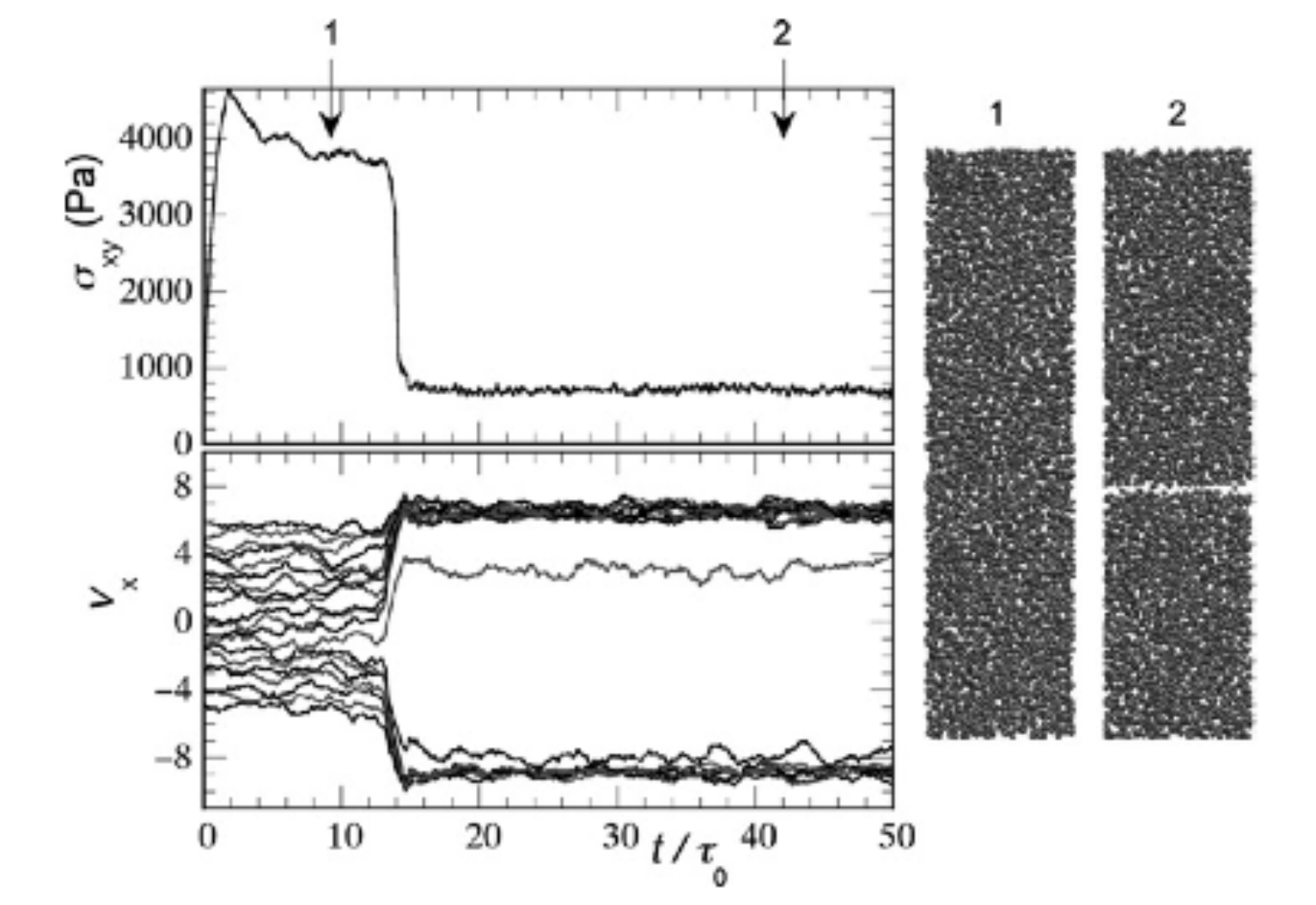}
\caption{ From [\cite{Sprakel2009}]: Transient shear stress response in a shear start-up run. Bottom panel shows the corresponding evolution of the velocity profile, in which each line represents the local fluid velocity $(v_x(y, t))$ at a given position in the gradient direction. Snapshots of the simulation box (flow from left-right with a velocity gradient from top-bottom) illustrate the homogeneous initial configuration (left) and the final fractured state (right).} 
\label{figure_simulation1}       
\end{figure}

Responsive particle dynamics simulations [\cite{Sprakel2009}] have also shown shear-induced  fractures (Figure \ref{figure_simulation1}) in transient polymer networks. Moreover these simulations have revealed a transition from shear banding to fracture upon increasing the overall polymer concentration.   and emphasize the difficulty to define an unambiguous criterion to discriminate between banding and fracture. In the literature, banding is usually defined as situations where the velocity profile in the gradient direction is continuous but kinked, whereas it is discontinuous at the plane of fracture. It is worth mentioning than the the issue of velocity change within the gap is in fact more complicated [\cite{Manneville2008}] in addition to "classic banding" coexistence of yielded and unyielded regions with finite and zero rates/velocities have been observed in  systems like colloidal glasses [\cite{Besseling2010}]. Unfortunately the spatial accuracy is typically limited to the micrometer range and cannot allow to distinguish between narrow high shear bands and fracture planes. However the authors of [\cite{Sprakel2009}] show that a discontinuity in the variation of the first normal stresses difference $N_1$ with the shear-rate is an unambiguous signature of a fracture, since normal forces should be largely reduced as all connections between the two shear bands across a fracture plane are broken. The discontinuity of $ N_1$  at the threshold  has been observed experimentally  for fracturing  transient networks [\cite{Tabuteau2009,Tixier2010}], in contrast with shear-banding regimes [\cite{Tixier2010}], where $N_1$ increases with $\dot{\gamma}$ as expected theoretically [\cite{Spenley1993}].

 \subsubsection{Fracture at constant applied stress}
 Another approach to study the failure of gels is to apply a constant stress and to follow the resulting shear rate as a function of time [\cite{Skrzeszewska2010}]. Above the critical fracture stress $\sigma\sim G_0$, fracture occurs immediately as revealed  by a dramatic increase of the measured share rate. Below the critical fracture shear stress, a delayed time is observed before fracturing, a phenomenon well documented in solid materials [\cite{Zhurkov1965,Santucci2007}] and observed also in colloidal suspensions [\cite{Sprakel2011b,Divoux2010}],  wormlike micelles [\cite{Olsson2010}] and  soft elastomers [\cite{Bonn1998}]. The delay fracture time  $t_b$ varies significantly from one experiment to an other one indicating  the stochastic nature of the fracture mechanism. With decreasing stress, the average delay time increases $<t_b>$  rapidly.
Several theoretical interpretations  have been proposed, that will be reviewed in Section \ref{sec : theorie}). They predict different delay fracture times. However, the intrinsic stochastic nature of the fracture mechanism  implies a very large uncertainty of $<t_b>$ that does not allow to conclude definitively about the relevant theory.
\begin{figure}[!h]
\begin{center}
\includegraphics[width=0.7\textwidth]{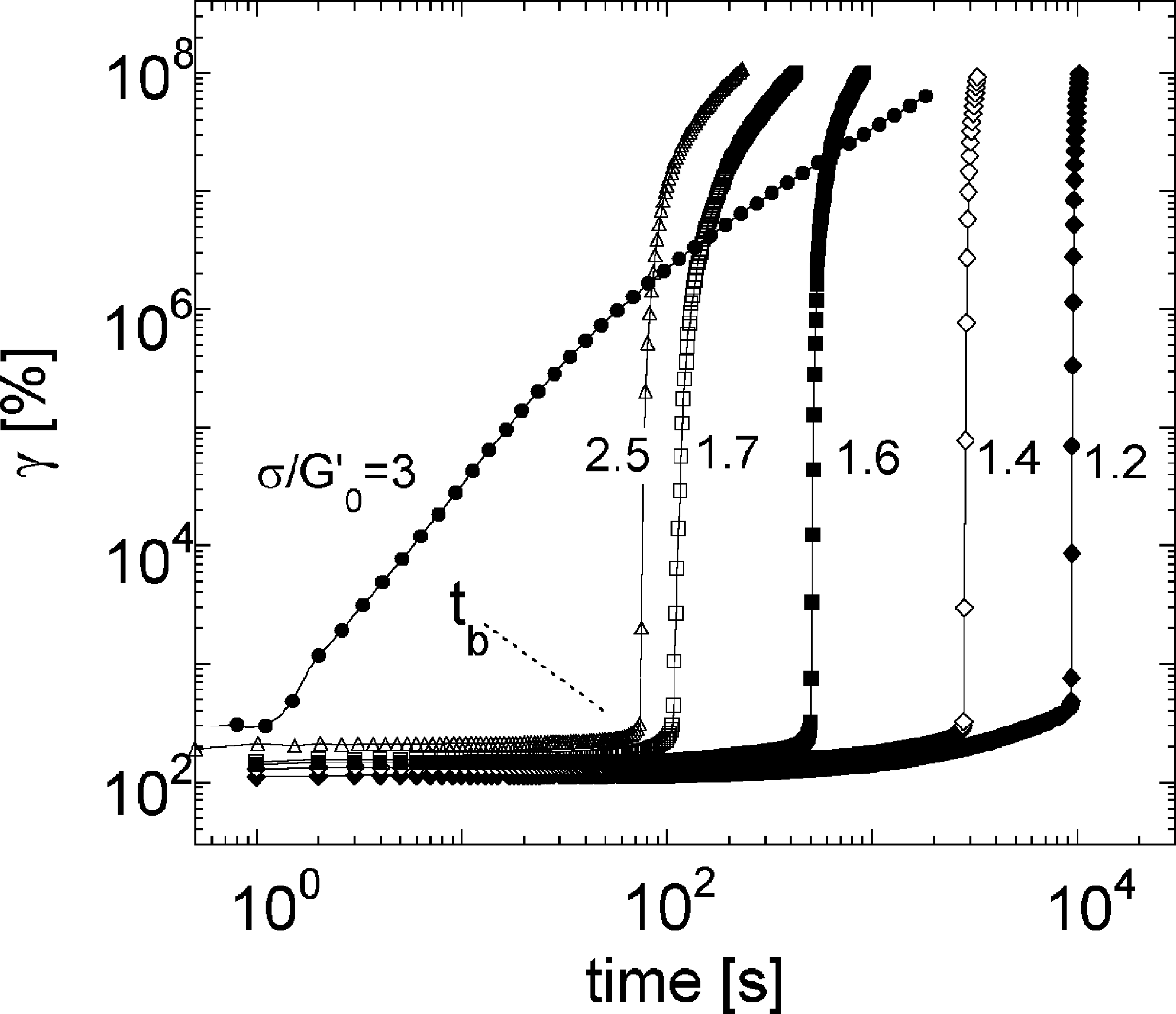}
\end{center}
\caption{From [\cite{Skrzeszewska2010}]: Delayed fracture of a transient polymer network formed by telechelic polypeptides, for different applied stresses. After a certain waiting time which varies from one experiment to another, the gel breaks, leading to a very rapid increase of the overall strain.} \label{fig : sprakelmacromolecules1}
\end{figure}

  \subsection{Fractures in Hele-Shaw cells} \label{sec: Hele-Shaw}
Fracture-like flow instabilities that arise when a fluid is injected into a Hele-Shaw cell filled with an aqueous solution of telechelic polymers has  been reported by several authors [\cite{Zhao1993,Ignes-Mullol1995,Vlad1999,Mora2010}].
   \begin{figure}
   \centering
  \includegraphics[width=12cm]{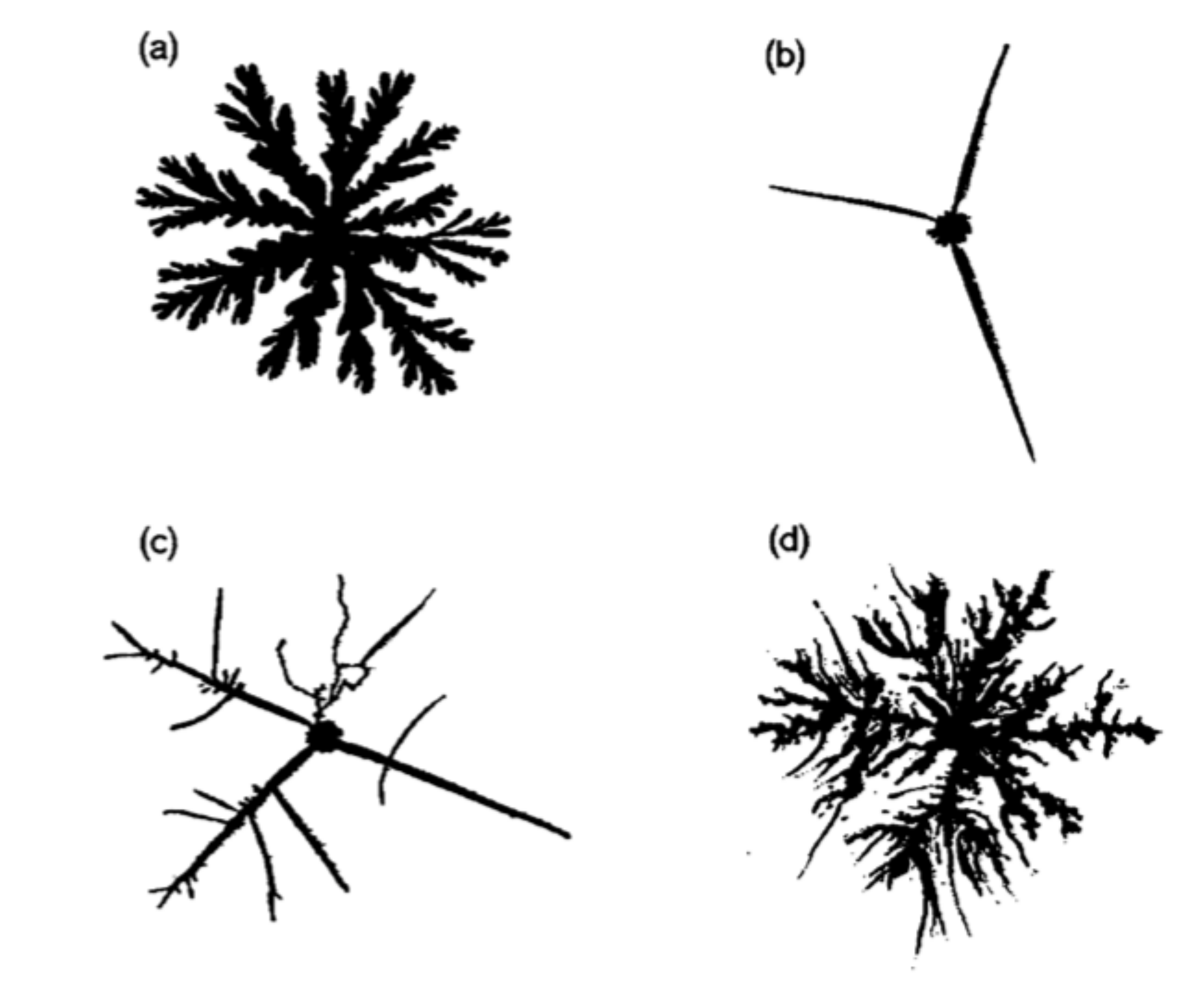}
\caption{ From [\cite{Zhao1993}]: Four typical patterns of a $2.5\%$ solution of hexadecyl end-capped polymers of molecular weight 50 700 at injection rates (a)$ Q = 1.0$ mL/min, (b) $ Q = 1.0 $mL/min, (c)  $Q = 5.0 $ mL/min, and (d)  $Q = 20$ mL/min. The crossover from the viscous-fingering pattern to the fracturing pattern is seen in (a) and (b).} 

\label{figure_HeleShaw}       
\end{figure}
 Zhao and Maher [\cite{Zhao1993}] showed using a radial Hele-Shaw cell,  that for transient networks  of associative polymers, there exists a threshold injection rate, below which the pattern instability is typically viscous fingering (Saffman-Taylor instability) [\cite{Bensimon1986}], and beyond which, the patterns resemble fracture patterns observed in brittle materials (Figure \ref{figure_HeleShaw}).  Note that analogous fracture patterns have been also observed in other classes of complex fluids like  clay suspensions [\cite{Lemaire1991}] or lyotropic lamellar phases [\cite{Greffier1998}]. The difference between a fingering pattern and a fracture pattern is drastic and can be quantified by calculating the mass fractal dimension of a given pattern: it drops drastically from $\sim 1.70$ to $\sim1.0$ near the transition threshold.  Interestingly, the fingering/fracture transition is not observed for  the corresponding entangled  homopolymers solutions, indicating  that the fingering-fracturing transition is a direct manifestation of an associating-network effect. A characteristic, Deborah number $ D_e=\tau_{thin}/\tau_f\sim(\tau Q)/(b^2L_{cell})$ can be defined, where $\tau_{thin}=1/\dot{\gamma_{thin}}$ is the inverse of the shear thinning rate of the transient network and {\em not} its  relaxation time $\tau$, $Q$ is the injection rate, $b$ the thickness of the cell and $L_{cell}$, some characteristic length scale of the radial cell.
The transition from fingering to fracture-like behavior appeared roughly at the same Deborah number under variation of polymer molecular weight or polymer concentration.

\begin{figure}
     \centering
  \includegraphics[width=12cm]{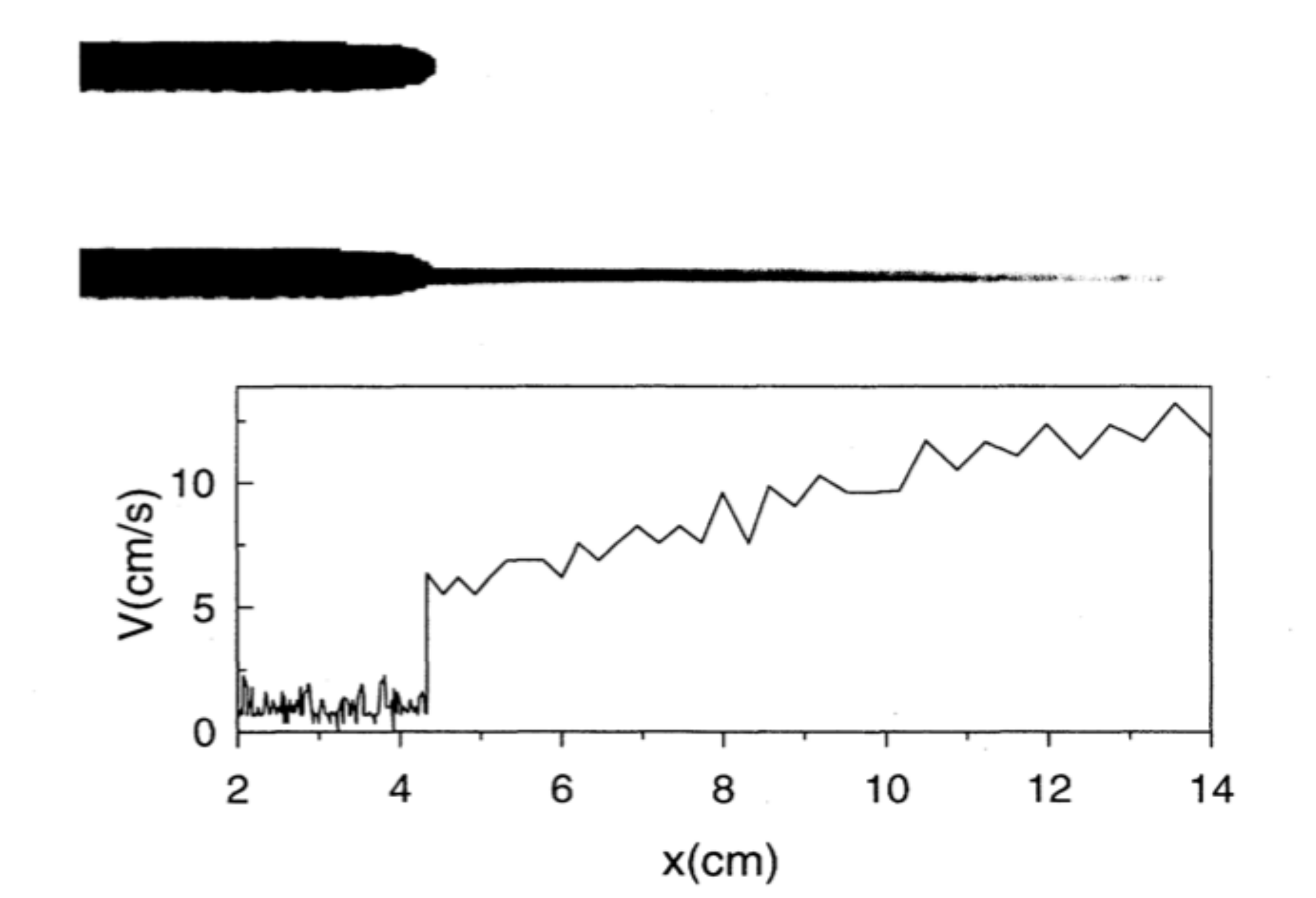}
\caption{ From [\cite{Ignes-Mullol1995}]: Fracture involving an associative polymer polymer of molecular weight 17400. \textbf{Top}: viscous finger at early times. The shape is that of a Saffman- Taylor finger. \textbf{Middle}: the fracture-like pattern emerges cleanly from the unperturbed finger. \textbf{Bottom}: tip velocity as a function of tip position during the growth).} 

\label{figure_meandering}       
\end{figure}

 Dynamics  fracture-like flows instabilities of associating polymers solutions has been also  characterized in a  rectangular Hell-shaw cell by measuring the crack  tip velocity [\cite{Ignes-Mullol1995}] or the tip mobility which relates the velocity of the tip to the pressure gradient [\cite{Vlad1999}]. The resulting pattern of the flow instability is highly dependent on the concentration, molecular weight or architecture of the associative polymer solutions. For low molecular weight polymer, the onset of fracture-like pattern evolution is  accompanied by an abrupt jump in the tip velocity followed by a constant low acceleration (Figure \ref{figure_meandering}). At high molecular weight, the transition is blurred by a meandering regime in which the crack tip meanders from side to side  and fluctuates with a characteristic frequency proportional proportional to the fluid injection rate.
 
\begin{figure}
    \centering
  \includegraphics[width=12cm]{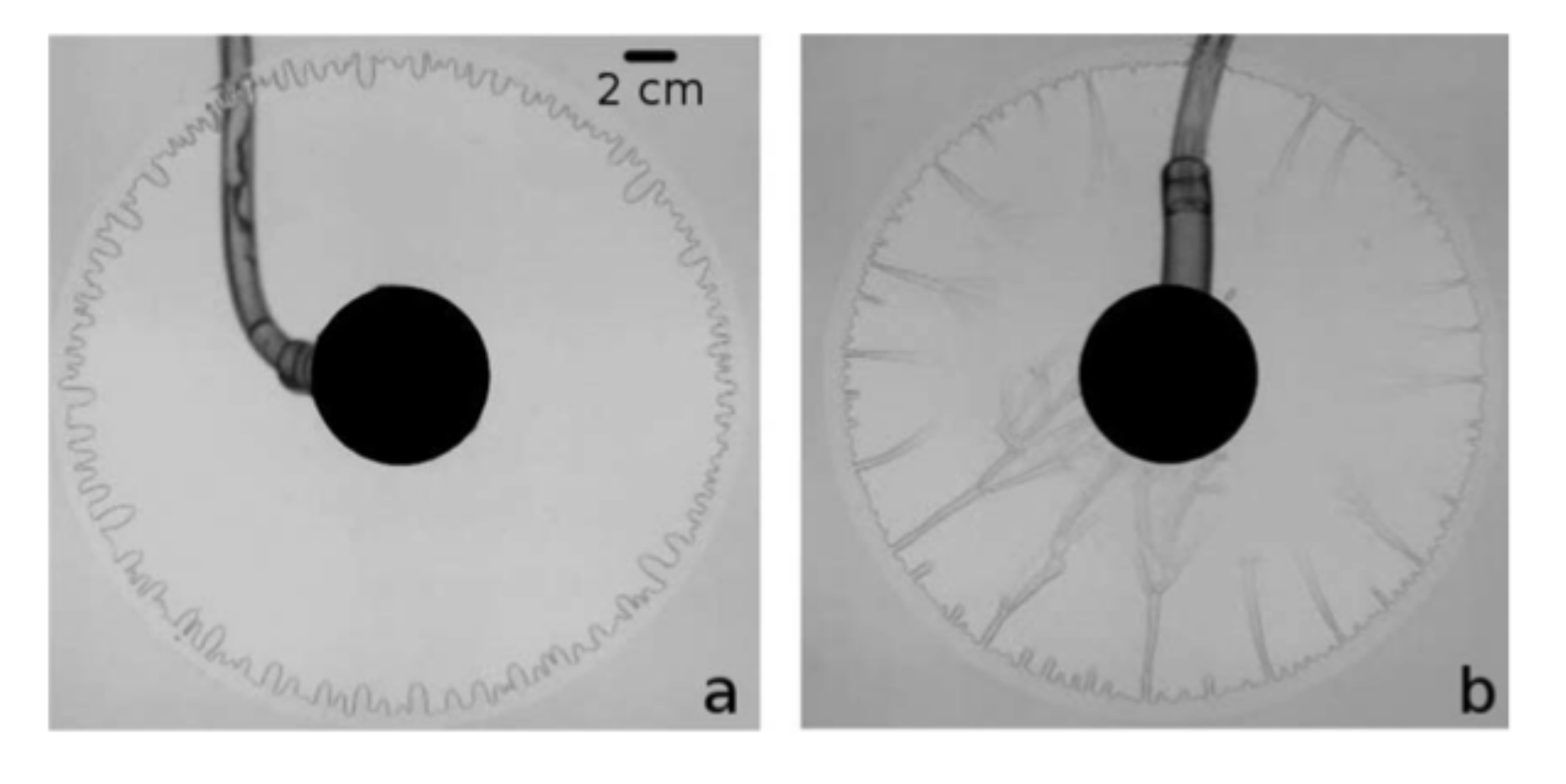}
\caption{ From [\cite{Mora2011}]: Aspiration of a Maxwell fluid in a radial Hele-Shaw cell: (a) Elastic pattern formation for  $\tilde{\tau}=5$, (b) Fracture-like pattern   for $\tilde{\tau}>10.$} 

\label{figure_Morapattern}       
\end{figure}

 Mora and Manna [\cite{Mora2009,Mora2010,Mora2012}] revisited both theoretically and  experimentally the Saffman-Taylor instability for  a model viscoelastic fluid, i.e. a upper-convected Maxwell fluid, a versatile experimental realization  of which consisting a transient network of oil in water microemulsion droplets droplets reversibly linked by telechelic polymers [\cite{Filali1999}]. They prove [\cite{Mora2010}] that a unique dimensionless parameter $\tilde{\tau}=\frac{b^2(-\nabla P)^{3/2}}{G_0\Gamma^{1/2}}$ ($b$ is the thickness of there Hele-Shaw cell, $\nabla P$  is the applied pressure gradient, $\Gamma$, the surface  tension of the network and $G_0$ is the zero shear modulus),  controls all elastic effects and particularly, the fracture transition. Note that $\tilde{\tau}$ does not depend on the fluid viscosity but on the elastic modulus. The theory describes the transition from viscous to elastic fingering instability  and shows that there is a divergence of the maximum growth ratio of the instability, which occurs for a finite value of the control parameter $\tilde{\tau}=10.2$. Experimentally (Figure \ref{figure_Morapattern}), the main results is the fracture-like pattern observed above a critical value $\tilde{\tau}^{exp}_c\simeq 10$ for every fluid, very close to the predicted theoretical value. So, the fracture is a direct consequence of the predicted blow up and the existence of the fracture-like pattens is a consequence of the elasticity of the complex fluid and takes place within the linear regime.

   \subsection{Fractures in extensional geometry} \label{sec : extensional}
     \subsubsection{Filament stretching rheometry}
   
\begin{figure}
   \centering
  \includegraphics[width=12cm]{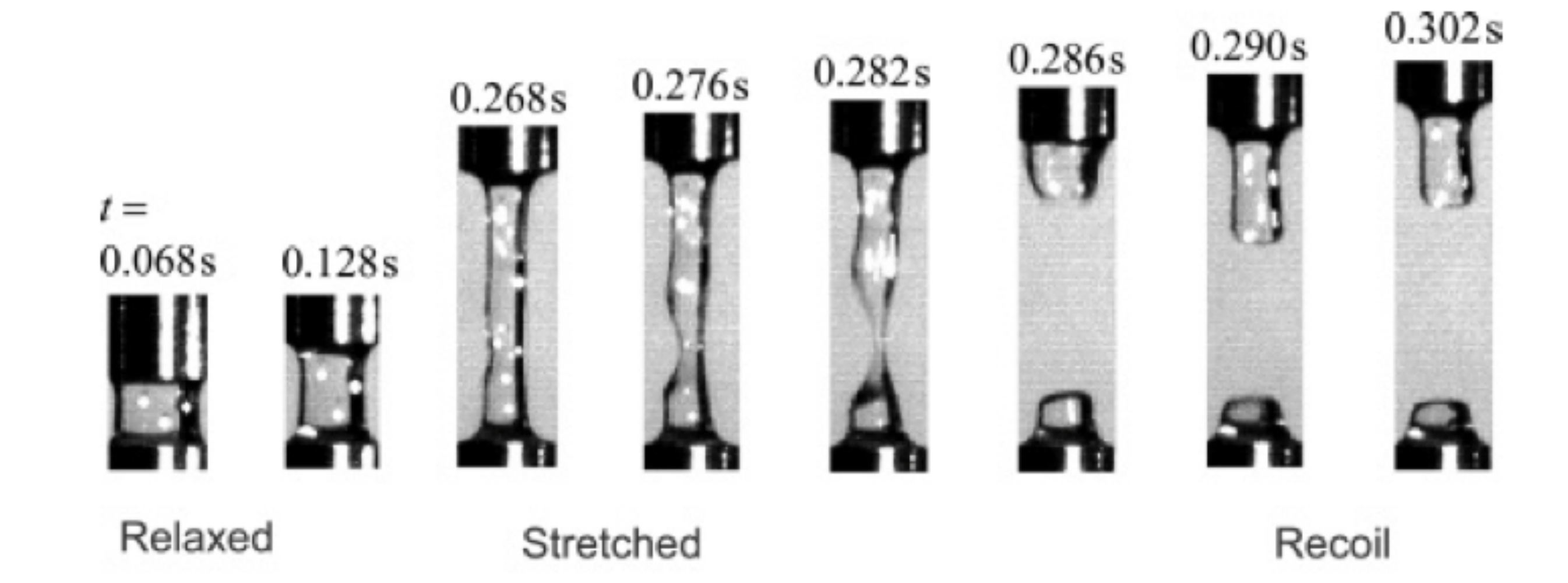}
\caption{ From [\cite{Tripathi2006}]:  Images showing rupture of 2.4wt \% HEUR22-2, for an applied strain rate $\dot{\epsilon}= 3 s^{-1}$. 
}
\label{figure_Tripathi}       
\end{figure}

Extensional  properties including fractures of associative polymer networks remain virtually unexplored until the development of filament stretching rheometry  [\cite{Anna2001,McKinley2002,Anna2008}]. Tripathi \textrm{et al.} [\cite{Tripathi2006}]
 investigated the non-linear extensional rheology of end-capped urethane coupled poly-(oxyethylene) (HEUR) using  a filament stretching rheometer. A  nearly cylindrical sample of transient network fills the gap between two rigid circular plates that are moved apart to a final separation with an exponentially  increasing separation profile $ L(t)=L_0\exp(\dot{\epsilon}t)$, where $\dot{\epsilon}$ is the strain rate, the evolution of the tensile force and mid-filament diameter being measured simultaneously. The authors  varied both the polymer concentration and the molecular weight of the polymer. They observed that, for sufficiently low polymer concentration (sufficiently low network strength), and at high strain rates and Deborah number (defined as $De= \tau \dot{\epsilon}$, where $\tau$ is the Maxwellian  relaxation time), a defect appears in the filament  and a tear propagates rapidly across the filament  leading to a complete rupture in two distinct domains followed by an  oscillatory damping elastic recoil  of the two pieces toward the two end planes  (see Figure \ref{figure_Tripathi}). The failure of the filament occurs before any significant necking has occurred. This type of rupture instability has also been observed in polymer melts [\cite{Joshi2004}] and entangled solutions of wormlike micelles [\cite{Rothstein2003,Bhardwaj2007a}]. At lower Deborah numbers, the fluid filament pinches off through a visco-capillary  or elasto-capillary thinning, and occurs after a considerable necking of the filament. These two different rupture scenarios  (elastic rupture versus necking and pinch-off) observed in transient networks and more generally in polymeric liquids are reminiscent of brittle versus ductile fracture behaviors [\cite{Tabuteau2008}] observed in solid materials,  reviewed in section \ref{sec : ductile}.

Tripathi\textit{ et al}  [\cite{Tripathi2006}] interpret the failure behavior  of the transient network filaments in the context of the Consid\`ere construction [\cite{Considere1885}], originally developed in solid mechanics to quantitatively predict the critical Hencky strain to failure, despite the doubts about its applicability to the prediction of the inception of necking in extensional flow [\cite{Petrie2009}]. The original  Consid\`ere construction is that, if the  graph of force against stretch ratio has a maximum, there is the possibility of two stretch ratios corresponding to the same force, which can be  associated with the  formation of a neck  in the solid sample, leading to the failure. Renardy [\cite{Renardy2004}] noted that fluid threads described by constitutive models in which the extensional viscosity passes through a maximum may undergo a purely elastic mode of necking failure in which surface tension plays no role. The observations of Tripathi et al of  filament necking and rupture processes observed experimentally appear to be connected to an instability resulting from saturation in the tensile stresses  at a critical deformation rate. It has been  associated with a maximum in the extensional viscosity as a function of rate of strain in the constitutive  rheological model used to described transient networks, so that this maximum of extensional  viscosity is claimed to be an analogue for liquids of the Consid\`ere criterion for solids. However, such an analogy  to interpret  the failure of transient networks  requires great care and at worst leads to confusion and misleading conclusions as developed in the review of Petrie [\cite{Petrie2009}], due to the use of too crude constitutive models for the complex fluids and a strong dependence on the details of the extensional flow. Recently, Fielding [\cite{Fielding2011}] finds, contrary to the Consid\`ere criterion, the onset of instability to relate closely to the onset of downward curvature in the time (and so strain) evolution of the $z$ component of the molecular strain, for extension along the $z$ axis.
     \subsubsection{ Fracture of pendant drops} \label{sec : pendant exp}
     The fall  and failure of  a viscous  fluid droplet from a faucet has been extensively studied both experimentally an d theoretically  (see the reviews by Eggers  and Villermaux [\cite{Eggers2008}] and Eggers [\cite{Eggers1997}]). However, for non-Newtonian fluids, such a surface free flow lead to highly non-linear fluid motion, in particular strong stretching that are usually non-well tested  by Non-Newtonian  constitutive laws, and few  quantitative experiments have been available.  Smolka and Belmonte [\cite{Smolka2003}] reported the filament dynamics and rupture  of  pendant drops of viscoelastic wormlike micellar fluids: the rupture is consecutive of a necking and pinching off instability followed by a partial retract of the free filament ends, is also reminiscent of a ductile fracture behavior observed at low Deborah numbers in filament stretching rheometry experiments performed on transient networks [\cite{Tripathi2006}]. 
     
\begin{figure}
\centering
  \includegraphics[width=8cm]{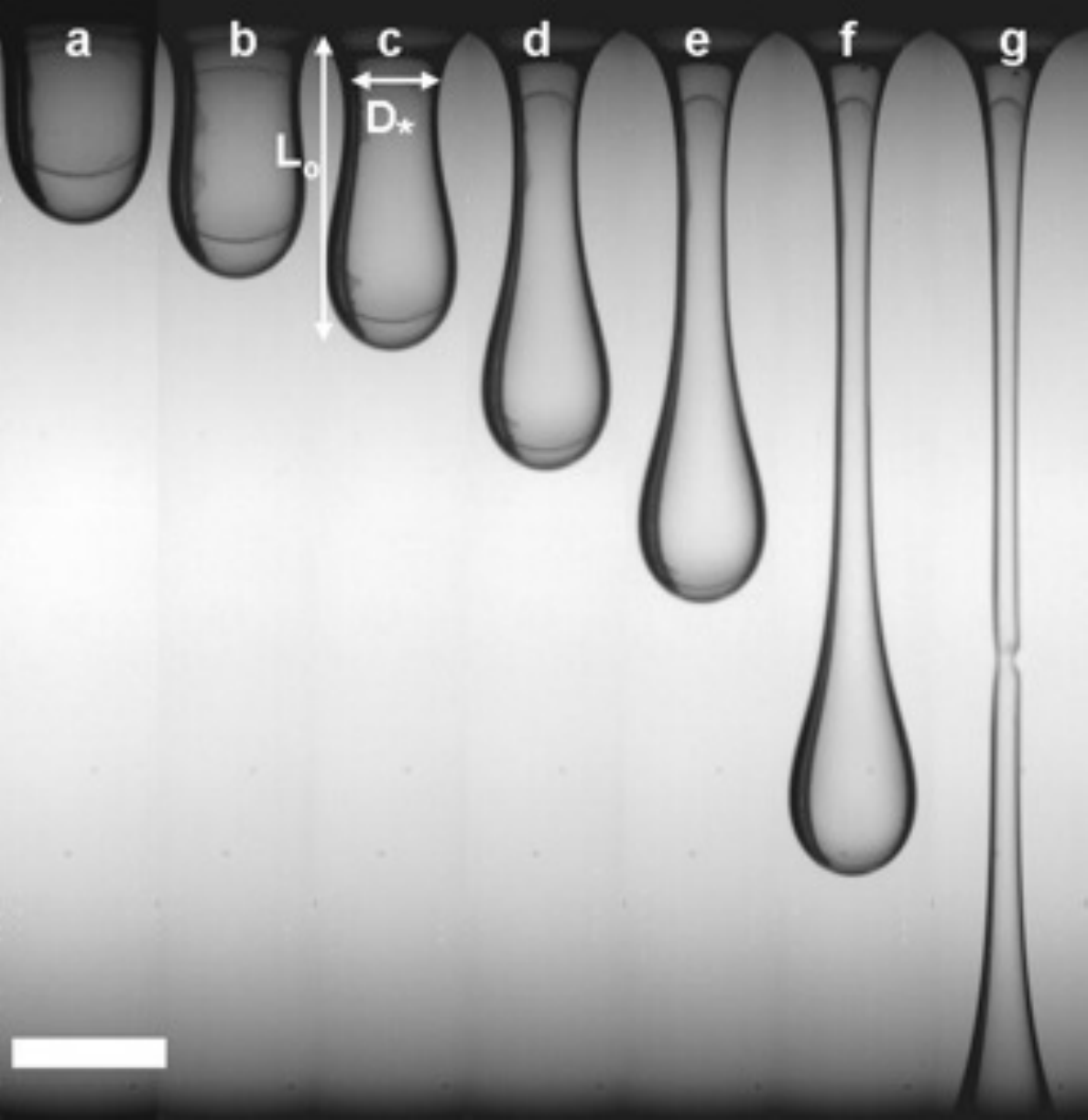}

\caption{ From [\cite{Tabuteau2011}]:  Sequence of images of the fall of a  drop of a transient network. Regime 1 (drop formation) corresponds to images (a) and (b) and regime 2 (viscous fall) corresponds to images (c) to (g). Image (g) corresponds to the beginning of the brittle fracture regime. At the beginning of regime 2, the radius of the filament is defined as $R* = D*/2$ and $L_0*$ is the length of the drop. The white scale bar corresponds to 2.5 mm.}
\label{figure_dropshape}       
\end{figure}

     The single report (to the best of our knowledge) of  fail and rupture of pendant drops of transient network has been done by  Tabuteau \textit{et al } [\cite{Tabuteau2009}]. The experimental system they used is a model network consisting of oil in water microemulsion droplets linked by telechelic polymers (described in section 2), which has the peculiarity to exhibit a pure linear Maxwell  behavior as long as it flows. A drop emerges at a constant rate of 2mL/h from a plastic tube of 2.6 mm diameter. The drop of fluid begins to fall when its weight exceeds the surface tension retaining force; a filament is formed and stretched by the failing drop until  a crack nucleates somewhere around the filament (Figure \ref{figure_dropshape}) and  then the fracture propagates perpendicularly to the filament. The total duration of the fall before the opening is typically on the order of 50 to 200 s depending on the sample, much larger than the relaxation time of the fluid ($\sim$ 1s), itself  much larger than the total duration of the rupture event ($\sim$ few ms). After that  the complete rupture  was  achieved, an elastic retraction  of the remaining pendant piece of the filament is observed.  Upon growing, the fracture exhibits a parabolic shape expected for an elastic solid breaking under tension [\cite{Greenwood1981,Barenblatt1962}]. The authors  measured the diameter of the drop where the fracture occurs and  weighted the mass of the falling part, so they can measure the rupture stress  $\sigma_c$  as a function of the shear modulus of the gel, that can be varied by varying the polymer concentration.
Moreover, the authors performed several measurements of the rupture stress of two samples with the same composition (oil mass fraction: 10\% and mean number of polymer  sticker per droplet: 6) and the same shear modulus $G_0= 1200$ Pa {\em but} with two different lengths of the hydrophobic alcane chains constituting the stickers of the telechelic polymers  (21 methyl groups and 18 methyl groups respectively). The relaxation time of the C$_{21}$ sample is five times larger than the relaxation time of the C$_{18}$ sample. The mean value of the rupture stress is  found to be identical for the two samples, with also the same relative standard deviation. This indicates that the rupture stress is  \textit{independent } of the binding energy.
     
   \begin{figure}
   \centering
  \includegraphics[width=10cm]{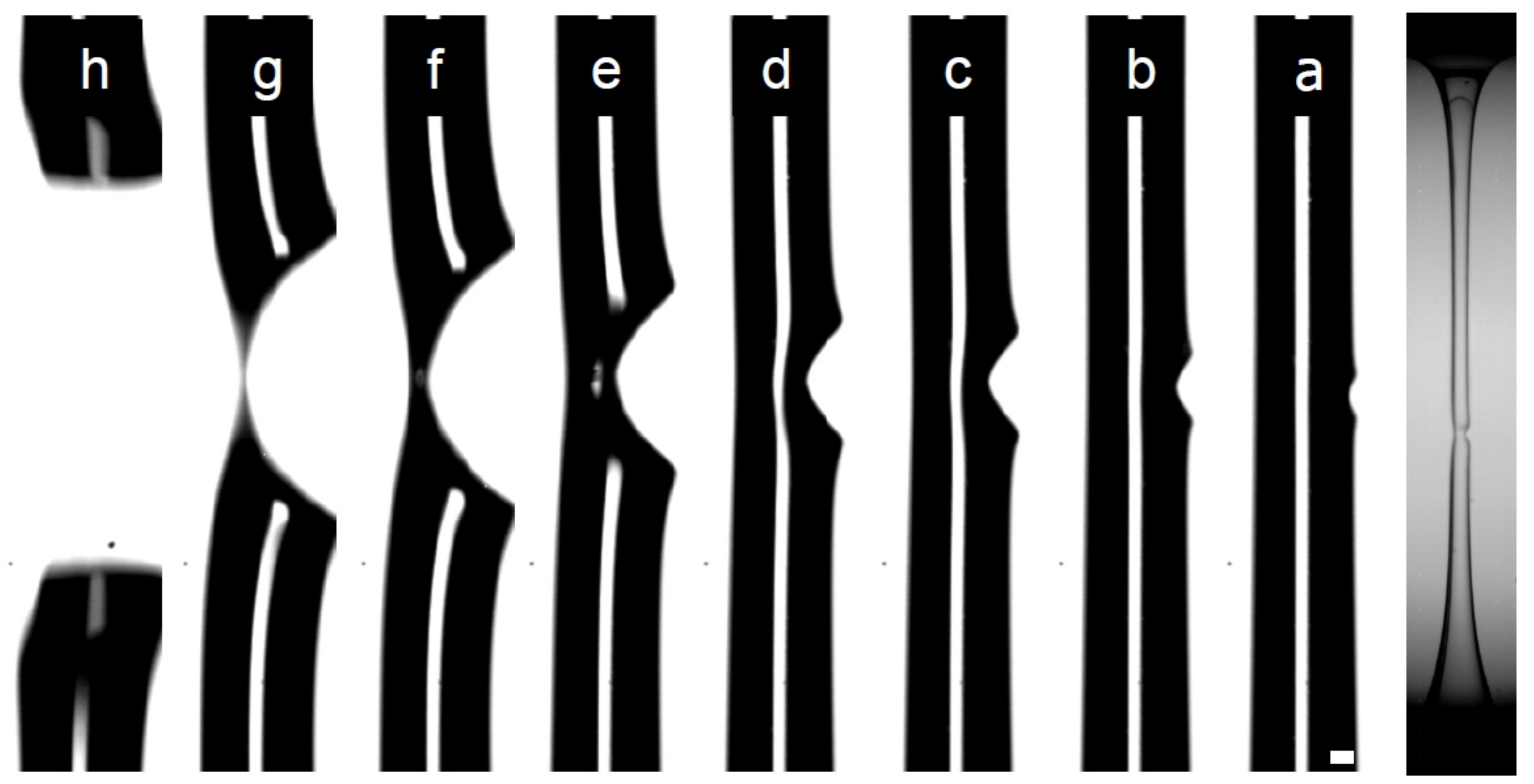}

\caption{ From [\cite{Tabuteau2011}];  Pictures of the propagation of a fracture across a pendant filament of a transient network from the right to the left. The time left to achieve complete fracture of the filament corresponding to each picture labelled with a letter is: a (8.50 ms), b (5.16 ms), c (2.67 ms), d (2.00 ms), e (1.00 ms), f (0.33 ms), g (0.17 ms) and h (0.ms). The last picture on the right shows almost all of the elongated drop, with the crack being well developed. The white scale bar corresponds to 0.1 mm.}
\label{figure_movie}       
\end{figure}   

\begin{figure}
   \centering
  \includegraphics[width=10cm]{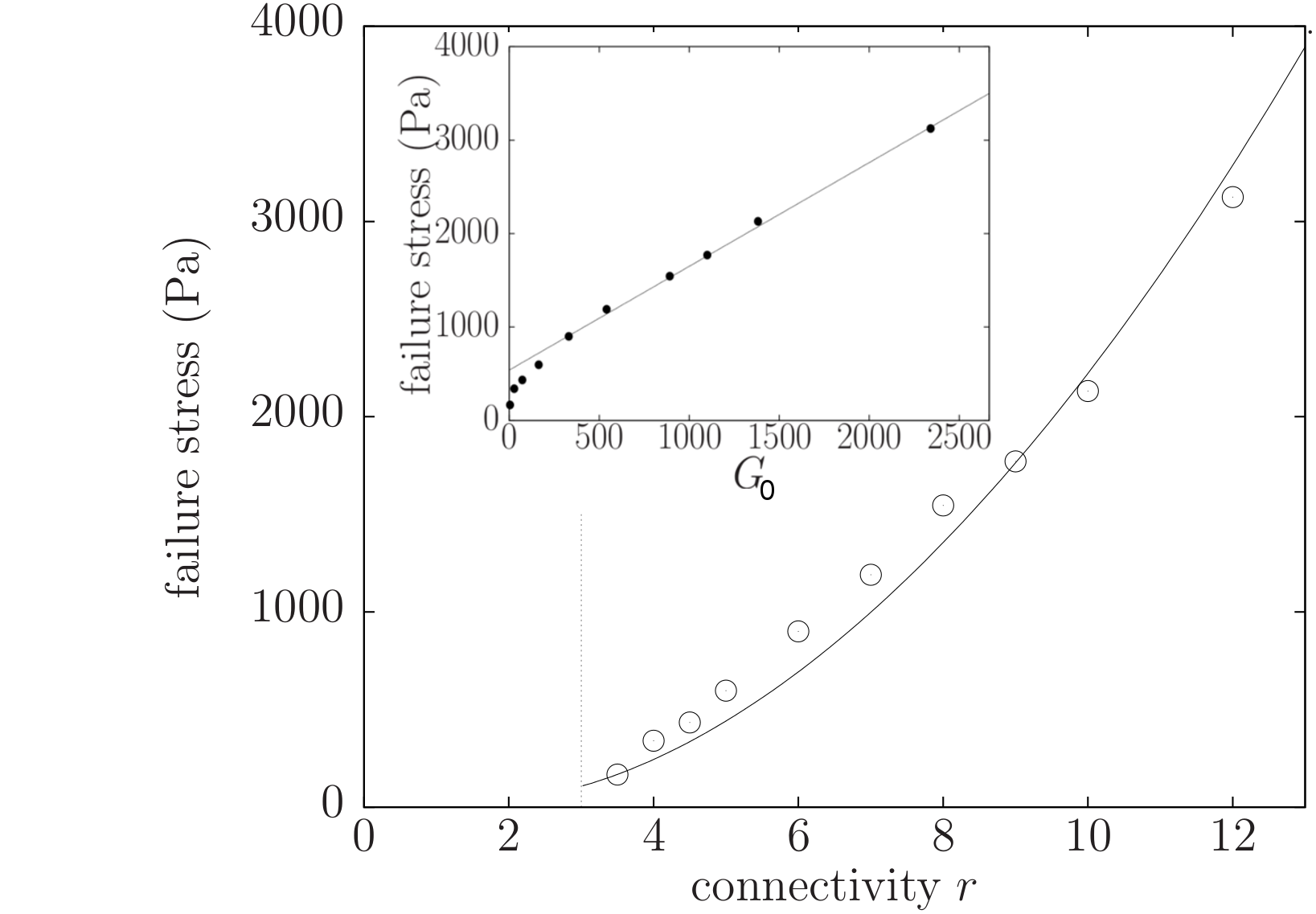}
\caption{ From [\cite{Tabuteau2009}]: Plot of the rupture stress coming from the pendant drop experiments ($\sigma_{b}$, circles) and the expected values coming from Eq. \ref{eqn : pomeau xi} ($\sigma_1$, line) as a function of the connectivity $r$.   The vertical dashed line corresponds to the percolation threshold ($r$=3). The inset gives the variation of the experimental rupture stress with the shear modulus, corresponding to the different connectivity. Note that the observed deviation of $\sigma_1$ from its linear dependence with the shear modulus $G_0$ observed at small $r$, originates from the vicinity of the percolation threshold [\cite{Filali1999}].}  
\label{figure_pomeau}       
\end{figure}

An other important point reported in [\cite{Tabuteau2009}] is that the rupture occurs in the linear regime $\sigma_c \approx G_0$, so that the critical rupture stress is close to shear modulus of the network (see Figure \ref{figure_pomeau}). The \textit{thermally activated fracture model} developed in section \ref{sec : thermally activated} allows a quantitative interpretation of these behaviors. One of the key points to understand this mechanism of fracture is the bond reversibility and the corresponding relevant ultra low interfacial energy needed to nucleate the crack. This interfacial tension results from the loss of conformational entropy of polymeric bonds near a crack interface and is typically on the order of few $\mu N m^{-1}$ [\cite{Sprakel2007}]. The origin of this  ultra low interfacial tension will be discussed in section \ref{sec : thermally activated}.


    \subsection{ Fracture propagation}
\begin{figure}
     \centering
  \includegraphics[width=10cm]{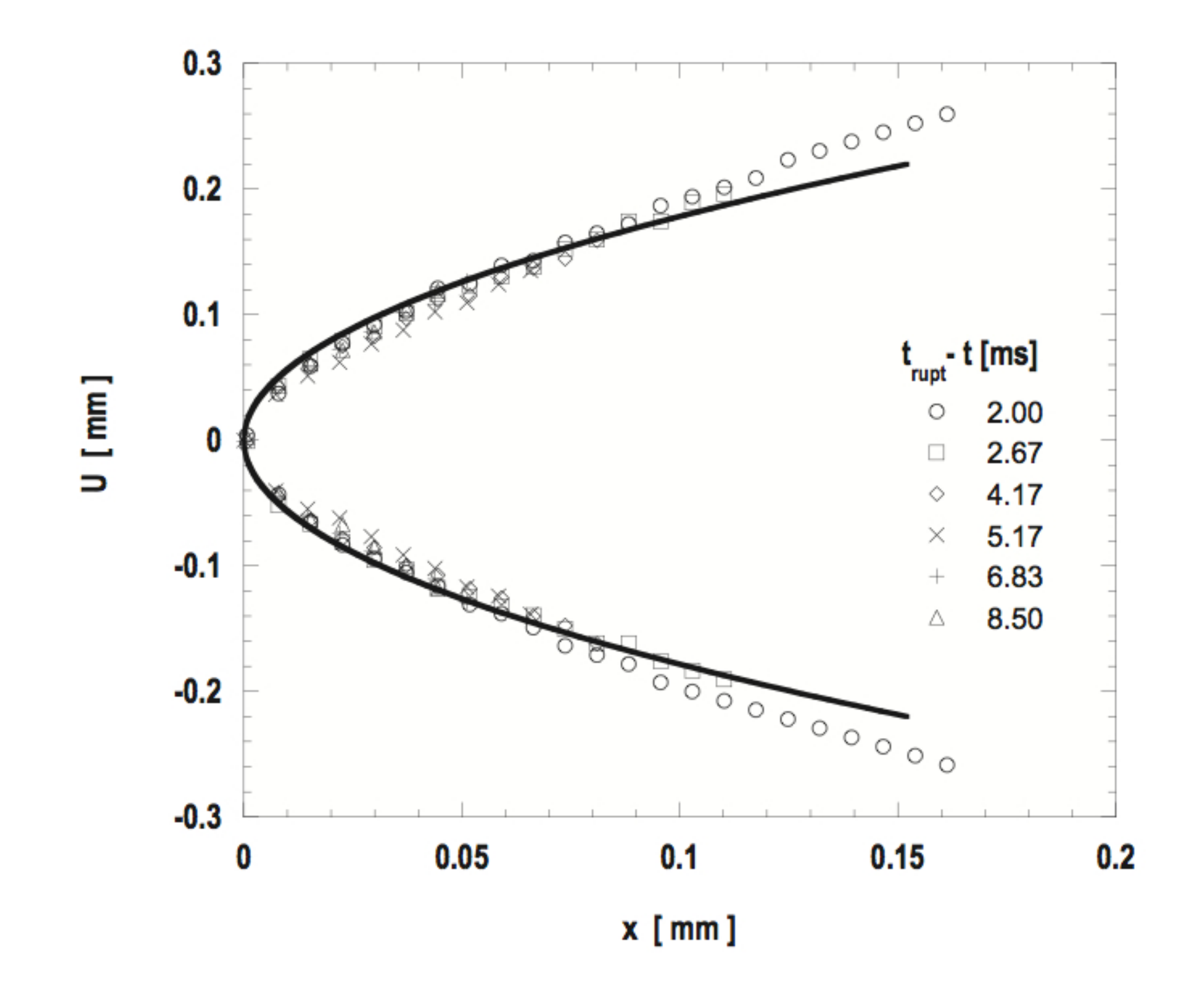}
\caption{ From [\cite{Tabuteau2011}]: Fracture profiles $u(x$) for different times before the break-up in the fracture moving frame, corresponding to the pictures of  figure \ref{figure_movie}. The black line is a parabolic fit corresponding to $u(x)=a\sqrt(x)$ with $J=\pi G_0 a^2/4$ and $J =2\Gamma_s$. We report only the profile for $L < 0.1D_0$ }
\label{figure_profiles}       
\end{figure}

    The propagation of fractures in transient networks is less documented [\cite{Ignes-Mullol1995,Vlad1999,Tabuteau2011}]  than in  permanent  gels, such as   covalent hydrogels [\cite{Tanaka2000,Livne2007,Livne2005}], biopolymer  hydrogels (gelatin [\cite{Baumberger2006,Baumberger2006b,Baumberger2010}]  or alginates [\cite{Baumberger2010b,Baumberger2009}]) and permanent  self-assembled triblock copolymer gels [\cite{Seitz2009}]. If one excepts some reports  on dynamics fracture of associating polymers in Hele-Shaw cells   [\cite{Ignes-Mullol1995,Vlad1999}]  already reviewed in section \ref{sec: Hele-Shaw}, the systematic analysis of morphology and propagation of a crack in a transient network   has been  reported very recently in a single publication [\cite{Tabuteau2011}]. The fracture propagation arising in pendant drop experiments  has been tracked by high speed velocimetry [\cite{Tabuteau2011}]. This configuration allows for an excellent reproducibility of the fracture initiation and propagation, as well as a pure elongational stress condition due to the lack of contact with solid interfaces near the fracture region. The fracture of the fluid happens in two steps. The fracture initiation step, reviewed  in the previous paragraph and reported in [\cite{Tabuteau2009}] was shown to be governed by the thermally activated nucleation of a critical crack in the polymer network. The second step, consisting in the rapid propagation of a brittle fracture in the fluid, was analyzed in detail  and shown to be energetically governed by the surface tension of the solvent (oil- in-water droplet microemulsion). Indeed, from the beginning of the propagation up to the complete fracture (i.e., when the sample is separated in two parts) the fracture profile exhibits a parabolic shape (Figure \ref{figure_movie})  as expected for an elastic solid breaking under tension [\cite{Irwin1957}]. These observations were confirmed by quantitative analysis of the fracture profile $u(x)$ on the crack propagation across the sample (Figure \ref{figure_profiles}).     
  
\begin{figure}
     \centering
  \includegraphics[width=8cm]{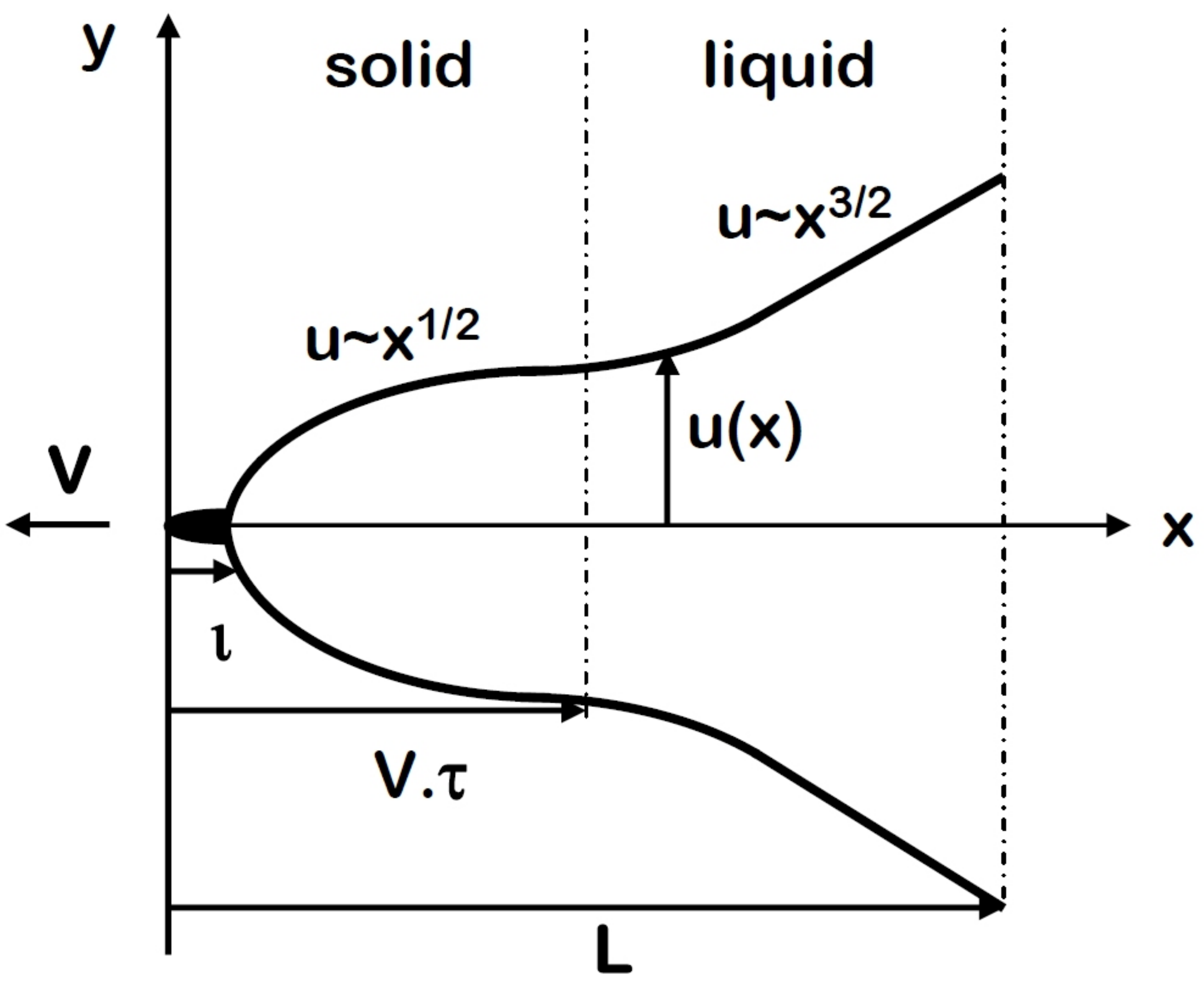}
\caption{ From [\cite{Tabuteau2011}]: A schematic representation of the space and time scales associated to a crack of length $L$ moving with velocity $V$ in a Maxwell fluid of relaxation time $\tau$ according to the trumpet model[\cite{deGennes1996}]. A small microscopic nonlinear zone of length $l$ is represented in black. The behavior of the material is solid like at scales smaller than $V\tau$, then fluid like at larger scales, and accordingly  the fracture profile scales differently with the  distance to the crack's tip in the two different regions. }
\label{figure_trumpet}       
\end{figure}

 Interestingly, the authors show that,  because of the low value of the elastic modulus of the transient network, and as in soft solids [\cite{Seitz2009}], the crack tip region exhibits very large deformations that require the use of finite elastic theories [\cite{Hui2003}] for a proper description. The non-linear character of the problem requires the fracture energy to be estimated by the J-integral method [\cite{Rice1968}] and the strain energy release ${\cal G}$ can be related to the curvature radius of the tip $R_{tip}$ through ${\cal G}\approx J=\pi G_0 R_{tip}/2$. The brittle character of the fracture propagation is confirmed by the fact that the  measured fracture energy is independent of crack  velocity $V$ and equal  to the Dupr\'e work, that is the reversible work needed to form two new air/gel surfaces $2\Gamma_s$ (dry fracture), without any bulk viscous dissipation.   According to the viscoelastic trumpet model of de Gennes [\cite{deGennes1996}] applied to a Maxwell fluid [\cite{Saulnier2004}], the absence of bulk viscous dissipation was justified by the short length $L$ of the crack (less than the filament diameter) in relation to the characteristic length $V\tau$ above which the viscous dissipation starts to become effective (Figure\ref{figure_trumpet}). An extension of the trumpet model to the hyper elastic case allows to confirm the weakness of the dissipated energy in terms of a new relevant length scale $ R_{tip} \sim {\cal G}/G_0$, of the same order of magnitude as the observed crack lengths. In agreement with this interpretation, the measured crack opening profiles present a constant parabolic shape.
 
\begin{figure}
\centering
\includegraphics[width=6cm]{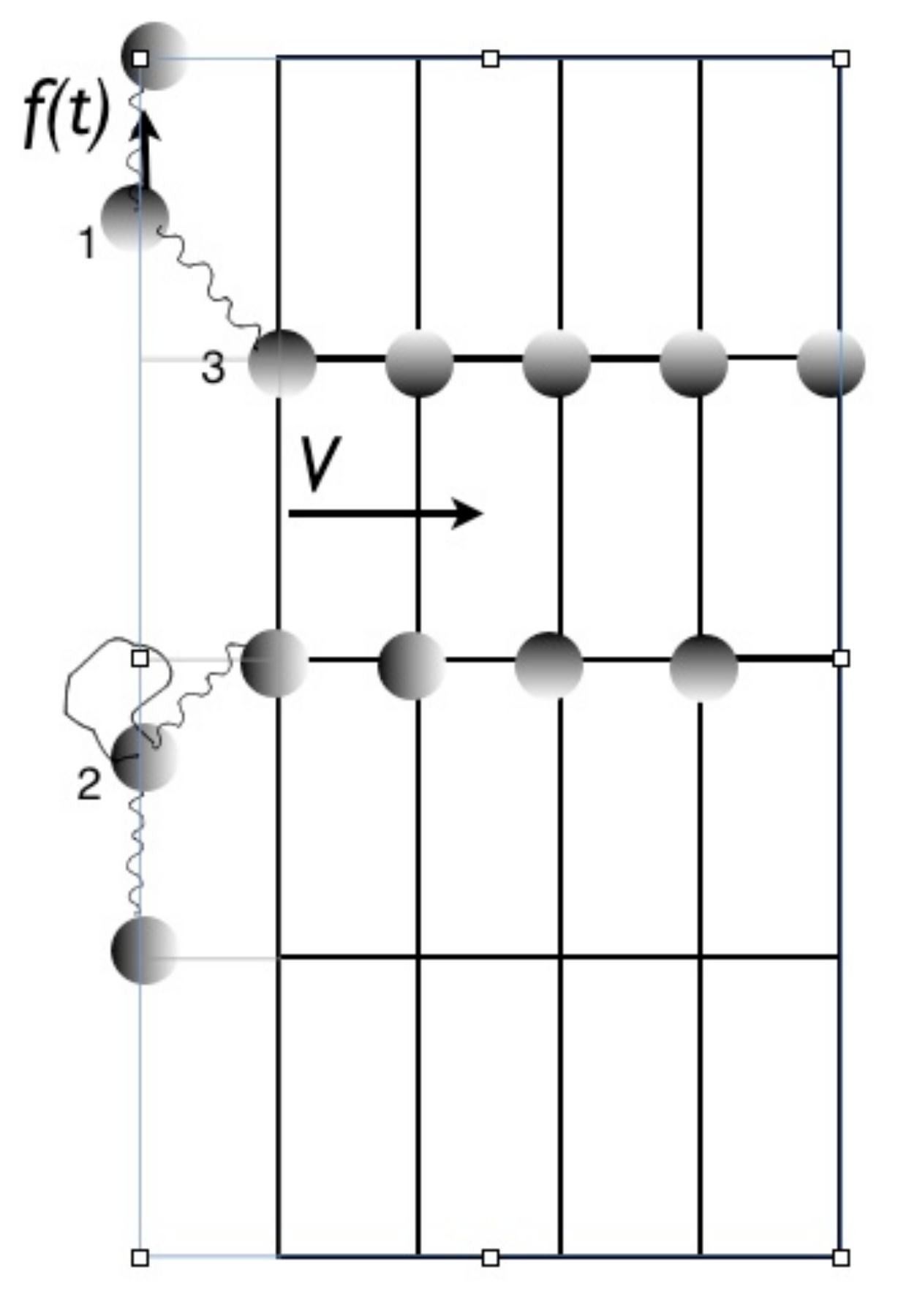}
\caption{From Ref [\cite{Tabuteau2011}]: Cartoon of the viscous relaxation mechanism of a bead at the tip of the fracture.  The polymer bridge between beads (1) and (2) just debonded, forming a loop on bead (2). Bead  (1)  thus experiences  the spring-back force $f(t)$ due to the gel under tension;  this will lead to an increasing extra tension on bead (3) and crack propagation at velocity $V$.}
\label{figure_friction}
\end{figure}

Despite the brittle character of the propagation, the authors report  a slow velocity of crack propagation $V\approx $ mm/s $\ll c_R$, $c_R\approx\sqrt{G_0/\rho}$ on the order of 1m/s, being the speed of sound in the medium, expected for a brittle solid. The authors propose a microscopic model for thee scaling of the velocity $V$ of the fracture to be governed by the time scales of the local crack tip debonding processes (Figure \ref{figure_friction}), and namely by the elastic relaxation time of the oil droplets after a debonding event under the action of the viscous drag of the solvent (Reynolds number $R_e\ll1$). This leads to an upper limit  for the estimation of the crack velocity $V\lesssim 0.25\frac{G_0 \xi^2}{2\pi\eta_Wb}$, where $G_0$ is the shear modulus of the gel, $b$ the radius of the droplet and $\xi$ the mesh size and $\eta_W$ is the viscosity of water. The control of the crack velocity by network/solvent friction has been already observed  by Baumberger \textit{et al.} [\cite{Baumberger2006,Baumberger2006b}] in the viscoplastic fracture dynamics of thermoreversible gels.

\section{Brittle vs Ductile fractures} \label{sec : ductile}
    Ductility versus brittleness characterizes the nature of fracture mechanism in solids under stresses when irreversible damages take place. Brittle solids suddenly break in the elastic regime above a critical stress, whereas ductile solids plastically deform, before fracturing. Ductility implies that some specific mechanism exists, allowing local rearrangements and creeping while the micro-structure is on the average preserved. Speaking of brittle fractures for viscoelastic fluids seems a priori paradoxical since, like all liquids, they accommodate arbitrarily large deformations after a finite relaxation time: in that sense they are infinitely ductile.  However, since the pioneering works of Vinogradov and coworkers in the 1970s and 1980s, it has been recognized that polymers melt can fail either in a ductile or a brittle mode, as discussed in details in [\cite{Wang2010,Boukani2009}]. The switch of the failure mode from yielding to elastic rupture is controlled  by the Rouse Weissenberg number.
 
\begin{figure}
\centering
\includegraphics[width=12cm]{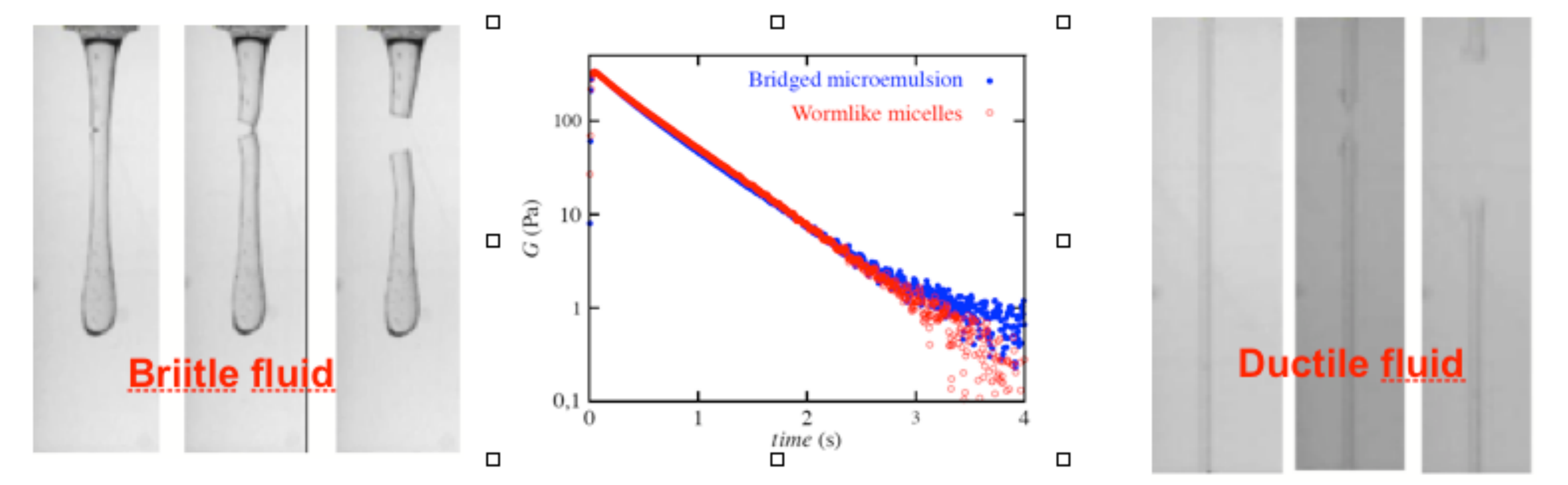}
\caption{ From Ref [\cite{Tixier2010}]: \textbf{Middle}: Shear stress relaxation function $G(t)$ for two model experimental systems of transient networks (bridged microemulsions and wormlike micelles). Both systems exhibit the same Maxwell behavior. \textbf{Left}: Fall of a droplet of the bridged microemulsion sample under the influence of gravity. Photographs taken at times (0, 10 ms, 20 ms) from the left to the right. The diameter of the filament is $\approx$ 5 mm. \textbf{Right}: Fall of a droplet of a wormlike micelles sample under the influence of gravity. Photographs taken at times (0, 66 ms, 67 ms) from the left to the right. The diameter of the filament is $\approx$2 mm.}
\label{figure_brittlevsductile}
\end{figure}

\begin{figure}
\centering
\includegraphics[width=12cm]{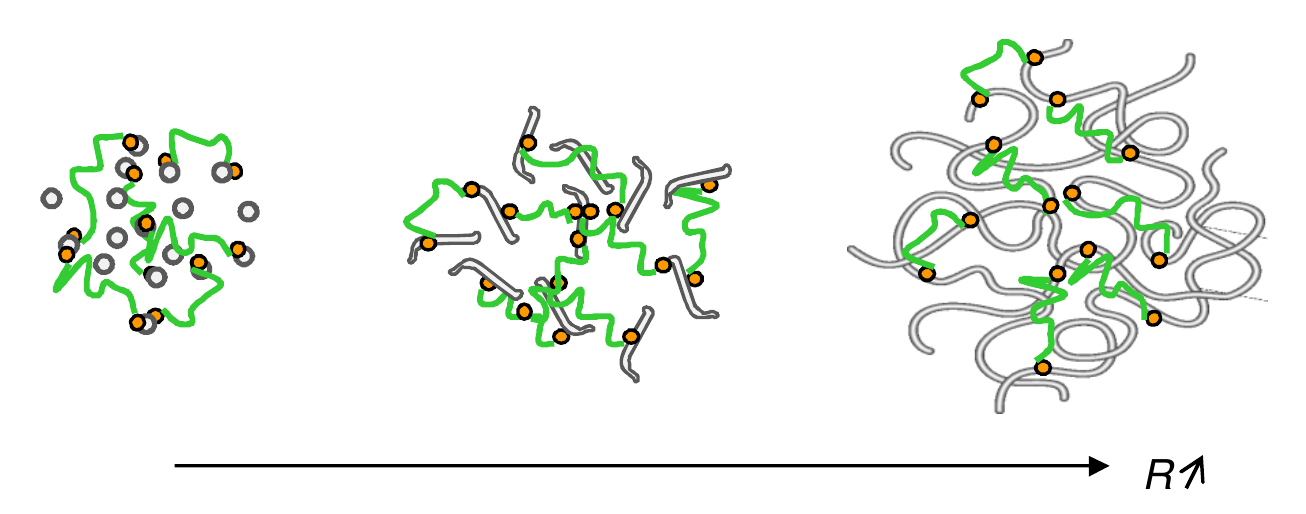}
\caption{ From Ref [\cite{Tixier2010}]: Schematic view of the structure of transient networks with nodes of tunable morphology. Upon variation of the co-surfactant to surfactant molar ratio of the micelles, the morphology of the nodes evolves from spherical (left) to rodlike (middle) to wormlike (right).}
\label{figure_tunablemicelles}
\end{figure}

   The  observation of fluid fractures in extensional geometry clearly reveals two  distinct rupture scenari  (elastic rupture versus necking and pinch-off) observed in viscoelastic fluids  and reviewed in section \ref{sec : extensional}, that are reminiscent of brittle versus ductile fracture behaviors observed in solid material. Interestingly, two transient networks (one consisting of cross-linked polymers the second of entangled living polymers [\cite{Dreiss2007}], the other  having the same simple  Maxwell linear rheological behavior exhibit the two distinct different fracture behaviors as shown in pendant drop experiments [\cite{Tabuteau2008}] (Figure \ref{figure_brittlevsductile}). Intuitively, the fracture should occur when the material does not have enough time to relax, i.e. when the flow time scale is faster than the relaxation time. This condition is equivalent to the condition that the Deborah number is larger than 1 [\cite{Gladden2007}]. However, as shown in Figure \ref{figure_brittlevsductile} and in [\cite{Tabuteau2008}], this appealing simple picture is a bit too naive. Bhardwaj \textit{et al} [\cite{Bhardwaj2007a}] observed both types of fracture in solutions of entangled micelles and suggested that the smooth pinch-off mode observed for the more concentrated sample might be the results of structural modifications of the solution of entangled wormlike micelles. On the other hand, Sprakel \textit{et al} [\cite{Sprakel2009}] investigated flow failure modes for transient polymer networks using particle-based simulations and showed transition from shear banding to melt fracture upon changing the polymer concentration. 
    Tixier \textit{et al} [\cite{Tixier2010}] reported the existence of a new class of self-assembled transient networks made of surfactant micelles of tunable morphology (from spheres, to rodlike to wormlike, Figure \ref{figure_tunablemicelles}) reversibly linked by telechelic polymers, designed to achieve a comprehensive study of the brittle to ductile transition in self-assembled networks. Flow curves of the transient networks of tunable morphology suggest that one can associate three domains  as the  micelles growth with distinct failures modes: a brittle mode, an intermediate mode and finally a ductile/shear banding mode, as the micelles grow.
         
\begin{figure}
\centering
\includegraphics[width=11cm]{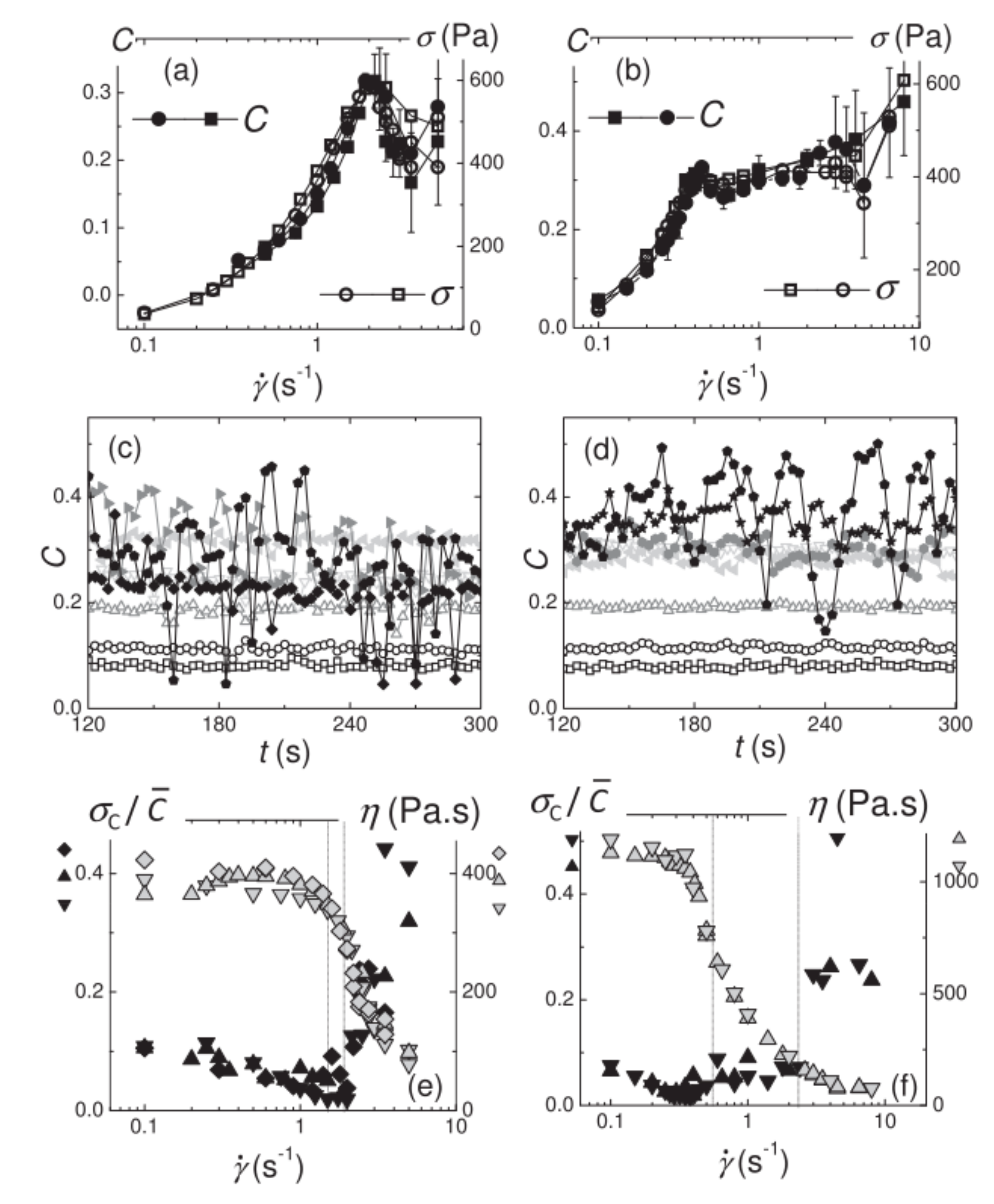}
\caption{ From Ref [\cite{Ramos2011}]: (a), (b): Shear stress (open symbols) and Contrast (full symbols) of the applied shear rate, $\dot{\gamma} $, in the steady state. The two sets of data in (a) and (b) correspond to two independent measurements. The bars for the contrast data correspond to the standard deviation. (c),(d): Time evolution of the contrast in the steady state as a function of $\dot{\gamma} $. (e),(f) Normalized fluctuations of the contrast (black symbols) and viscosity (gray symbols) as a function of  $\dot{\gamma} $. (a), (c), (e): brittle fracture behavior; (b), (d), (f) ductile fracture behavior.}
\label{figure_XRay}
\end{figure}

Coupling rheology  and  time-resolved structural measurements using synchrotron radiation [\cite{Ramos2011}], Ramos and Ligoure show that the bridged micelles can align under shear and proposed the emergence of strong fluctuations of the degree of alignment  of the micelles as a structural probe of a fracture process under shear (Figure\ref{figure_XRay}). The transition from a regime where the  fluctuations of the  are weak to a regime where the fluctuations are strong is rather sharp and enables  to define a critical shear rate for the onset of strong fluctuations $\dot{\gamma}_{fracture}$. By analogy with solid materials, and mapping the strain in solids to the shear rate of viscoelastic fluids under shear, the authors  therefore propose to compare the shear rate for the onset of fracture  $\dot{\gamma}_{fracture}$ and the shear rate for the onset of non-linearity  $\dot{\gamma}_{thin}$. They define $\Delta$ as the normalized difference between the two critical shear rates $\Delta=( \dot{\gamma}_{fracture} -\dot{\gamma}_{thin})/\dot{\gamma}_{thin}$. This parameter $\Delta$ is analogous o the amount of plastic deformation accumulated before fracture in ductile materials. So, the critical shear rate for fracture is either very close to the shear rate at which the sample departs from its linear behavior (for ``brittle-like'' samples) or significantly larger than the shear rate at which the sample departs from its linear behavior (for ``ductile-like'') samples, and the ability of transient networks to exhibit a ductile-like behavior should be  related to its  ability to rearrange its structure at large scale  under flow, before relaxing its elastic energy.

    \section{ Theoretical approaches of crack nucleation in transient networks} \label{sec : theorie}
Experiments reported by different groups on various transient networks but with similar structures and microscopic dynamical properties lead to  several interpretations that do not seem at first glance compatible, leading to some confusion [\cite{Skrzeszewska2010,Berret2001,Tabuteau2008,Tabuteau2009,Tabuteau2011}]. Sections \ref{sec : thermally activated}, \ref{sec : activated rupture} and \ref{sec : spring} are devoted to theories for the nucleation of brittle fractures,  and to the confrontation of these theories with experimental results. Section \ref{sec : thermally activated} deals with the {\em thermally activated crack nucleation model} (TAC model) [\cite{Golubovic1991,Pomeau1992}]. In section \ref{sec : activated rupture} {\em  the activated bond rupture model } (ABR model) [\cite{Evans1997,Chaudhury1999,Skrzeszewska2010}] is introduced. The starting point of these two models is the Griffith theory for fracture in brittle materials [\cite{Griffith1921}], where a critical length for  crack growth is expressed as a function of the applied load (Section \ref{sec : griffith}). This length is reached by the thermal fluctuations in the TAC model, whereas it results of an acceleration in the rupture rate of the bonds near a crack tip in the ABR model. A third model, that we call {\em Self healing and activated bond rupture nucleation model} (SH-ABR  model)[\cite{Mora2011}] is introduced in section \ref{sec : spring}. It is based on the balance between cross-links disconnection and reconnection [\cite{Mora2011}] to control crack nucleation. Finally, the validity of these models is compared and discussed in section \ref{sec : validity}.

\subsection{Griffith Fracture Criterion}\label{sec : griffith}

\begin{figure}[h]
\begin{center}
\includegraphics[width=0.8\textwidth]{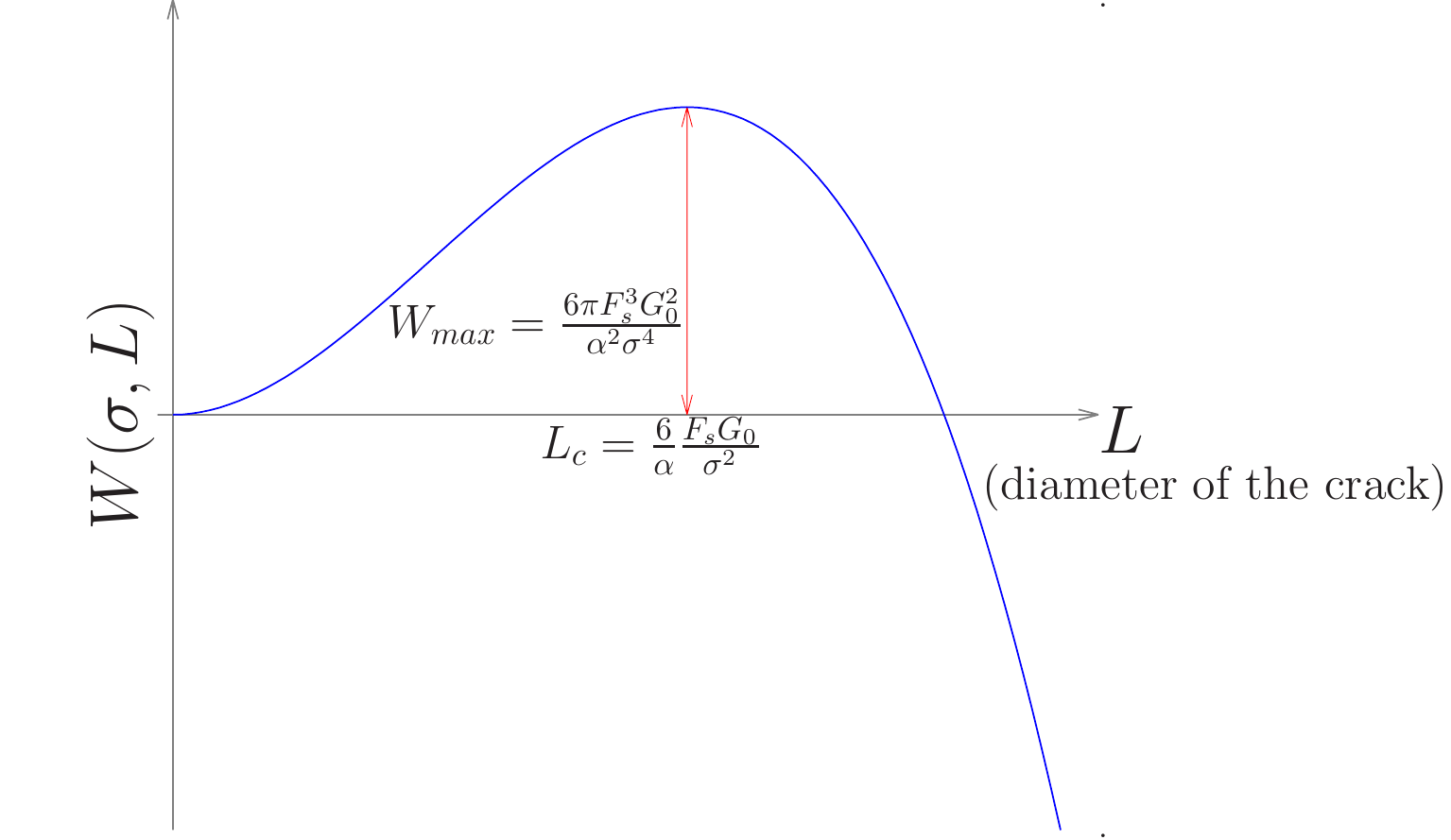}
\end{center}
\caption{Sketch of the Griffith's free energy of a crack as a function of  the crack length $L$. Note that this curve can be scanned in both directions (increasing or decreasing L) for self-healing materials, in contrast with  usual non-self-healing materials.} \label{fig : griffith} 
\end{figure}

The Griffith's theory for fracture in brittle materials [\cite{Griffith1921}] starts up with preexisting micro-cracks and considers the energy required for them to grow spontaneously under a given constant stress. An interfacial energy has to be paid to break more cohesive bonds, counterbalanced by the bulk elastic energy released by the opening of the crack. For a given stress $\sigma$ the Griffith  free energy cost  $W$ for a crack of size $L$  reads, in 3D:
\begin{equation}
W(\sigma,L) =\frac{\pi}{2} L^2 F_s-\alpha \frac{\pi L^3\sigma^2}{2G_0},
  \label{eqn : griffith} 		
\end{equation} 
with $L$  the size of a disc-like crack, $F_s$  the cohesive energy per unit area, $G_0$ is the shear modulus and  $\alpha \simeq 1$ a constant depending on geometrical factors. A typical example of the Griffith potential as a function of $L$ is shown on Figure \ref{fig : griffith}. The stress $\sigma$ being fixed, $W(\sigma,L)$ reaches a maximum:
\begin{equation}
W_{max}(\sigma) =\frac{6\pi}{\alpha^2}\frac{F_s^3G_0^2}{\sigma^4}
\label{eqn : maximum}        
\end{equation}
for a crack size (the so called Griffith length) 
\begin{equation}
L_c(\sigma)=\frac{6}{\alpha}\frac{F_sG_0}{\sigma^2}.
\label{eqn : griffith length}
\end{equation}
For a crack size larger than $L_c,  dW/d L<0$, leading to a catastrophic growth of the crack (fracture), in contrast with a crack size smaller than $L_c$. Thus, a stressed sample  appears to be in a metastable state as long as no crack with a size larger than $L_c$ nucleates  [\cite{Golubovic1991,Pomeau2002}].\\

\subsection{Thermally activated crack nucleation model (TAC model)} \label{sec : thermally activated}

A material submitted to a constant stress breaks after a given waiting time, that depends on the applied stress. Delayed fractures have been reported for a large variety of materials, ranging from 2d-microcrystals [\cite{Pauchard93,Pauchard1998}] to heterogeneous materials [\cite{Guarino1999,Guarino2001,Guarino02}] and  gels [\cite{Bonn1998}]. More recently, experimental evidences of delayed fractures have been reported in transient networks [\cite{Tabuteau2009,Skrzeszewska2010}] and a good agreement has been found with predictions of the TAC model [\cite{Pomeau1992}], whose starting point is the Griffith theory.
Several authors [\cite{Golubovic1991,Pomeau1992,Buchel1997,Pomeau2002}] have proposed that the  Griffith's  energy (Eq. \ref{eqn : griffith}) is an energy barrier of height $W_{max}(\sigma)$. Then, the waiting $t_1$ should follow the  Kramer's theory  [\cite{Hanggi1990}]:
\begin{equation}
\langle t_1 \rangle \propto exp\left( \frac{W_{max}(\sigma)}{k_BT}\right),
\label{eqn : arrhenius}
\end{equation}  
where $\langle.\rangle$ is for averaging over the stochastic distribution of fracture waiting times. The choice of an energy barrier implicitly assumes that a crack can reversibly explore states with different crack lengths between the initial and the critical ones.

A key characteristic of crack nucleation in transient networks is the intrinsic reversibility of the process. This directly comes from the transient nature of the junctions: new junctions can randomly appear between the two opposite sides of a crack. This is in fundamental contrast with what happens in  non-self-healing materials, where the crack length irreversibly increases, making Eq. \ref{eqn : arrhenius} irrelevant for solid materials. Recent models can overcome this specific difficulty of non-self-healing materials [\cite{Santucci2003,Santucci2007}].

\begin{figure}[!h]
\begin{center}
\includegraphics[width=0.52\textwidth]{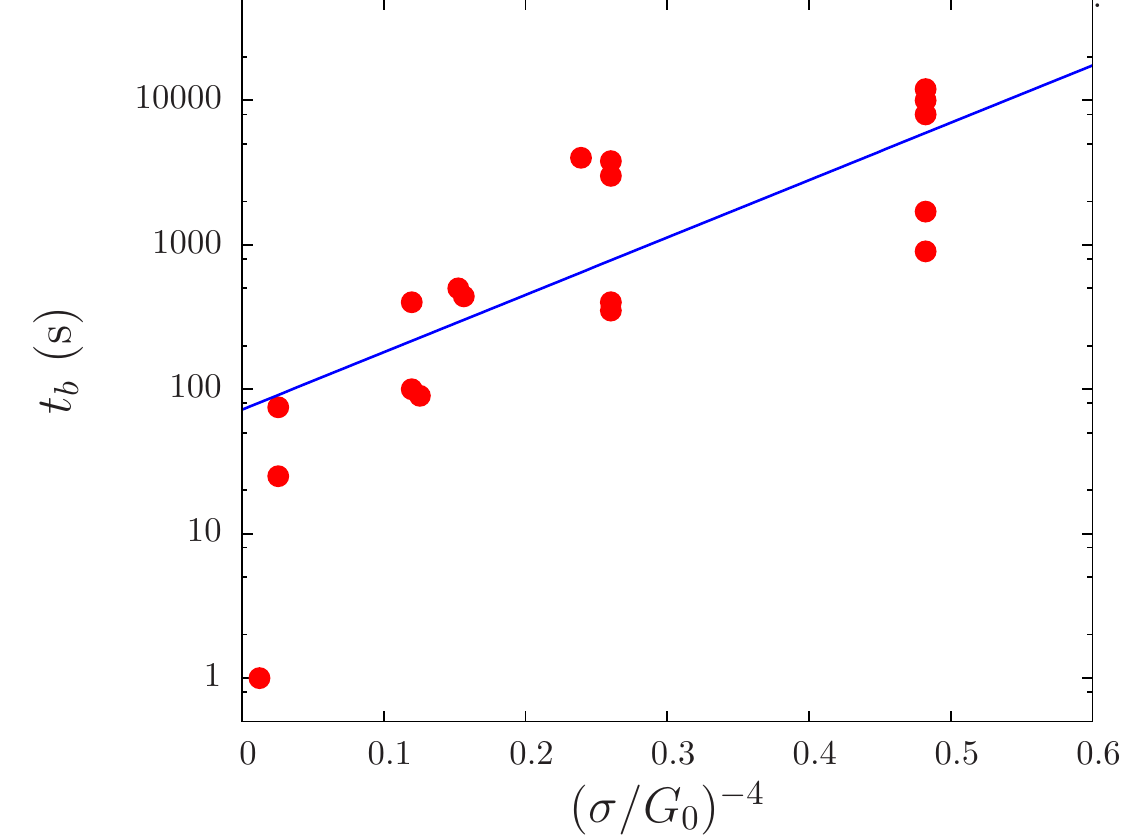}\hglue 0\textwidth \includegraphics[width=0.52\textwidth]{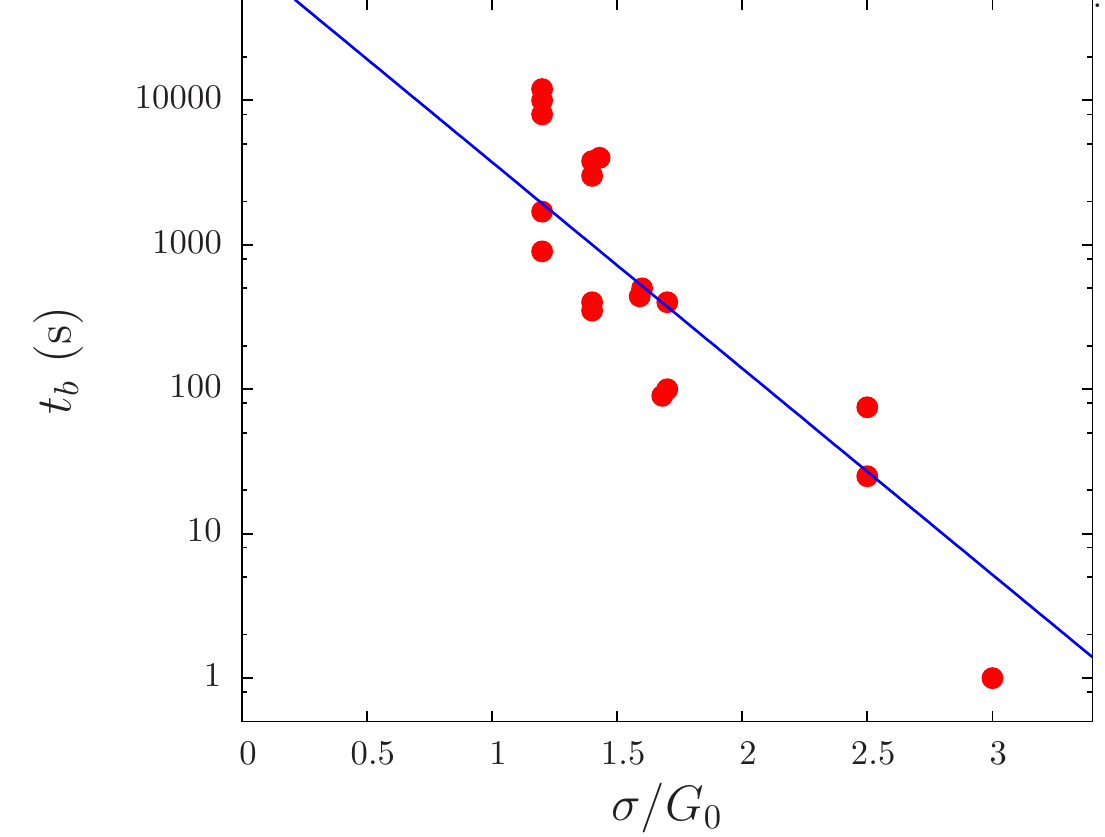}
\end{center}
\caption{Data from [\cite{Skrzeszewska2010}]: Measured fracture waiting time ($t_b$) as a function of the applied stress. {\bf Left}: $t_b$ as a function of $(G_0 / \sigma)^4$. The straight line represents the best fit for $t_0\exp\left(\left(\sigma_1/\sigma\right)^4 \right) $ ($\sigma_1/G_0= 1.74\pm 0.08$). {\bf Right}: representation in a semi-logarithmic scale. The straight line shows the best fit for an exponential decay $t_0\exp(-\sigma/\sigma_2)$ ($\sigma_2/G_0=3.3 \pm 0.6$).} \label{fig : sprakelmacromolecules2}
\end{figure}

A direct way to check predictions of the TA  model is to measure the average fracture waiting time for different values  the applied stress. Putting Eq. \ref{eqn : maximum} in Eq. \ref{eqn : arrhenius}, one obtains:
\begin{equation}
\langle t_1 \rangle \simeq \tau_1 \exp\left[\left(\frac{\sigma_1}{\sigma}\right)^4\right],
\label{eqn : waiting time}
\end{equation}
where the  prefactor $\tau_1$  varies much more gently as a function of $\sigma$, than the exponential for $\sigma \gg \sigma_1$. The variation of the fracture waiting time given by Eq. \ref{eqn : waiting time} is very stiff when the stress is close to the characteristic stress $\sigma_1$: the waiting time becomes very long as soon as the stress decreases below $\sigma_1$, and becomes extremely short if the stress increases beyond $\sigma_1$. This sharp variation makes difficult the experimental determination of the waiting time variations with respect to the applied stress: the average waiting time should vary over several decades to check Eq. \ref{eqn : waiting time}. Skrzeszewska et al. [\cite{Skrzeszewska2010}] have measured the fracture waiting time for transient polymer networks formed by telechelic polypeptides, with nodes made by the self association of three collagen-like triple helices [\cite{Skrzeszewska2009}]. The sample is put in a rheometer equipped with a Couette geometry, and a given constant stress is applied. The measured waiting time $t_b$ is defined as the elapsed duration between application of the stress and the instant when the measured shear rate jumps sharply by several decades (Figure \ref{fig : sprakelmacromolecules1}).  An important feature emerging from these experiments is the great dispersion for the measured values of the waiting times. This dispersion is inherent to the stochastic feature of the subcritical fracture. $t_b$ is plotted in Figure \ref{fig : sprakelmacromolecules2} for different values âthe imposed constant stress. According to available data, the standard deviation reached one decade. $t_b$ is plotted as a function of $(\sigma/G_0)^{4}$ (in semi-logscales, Figure \ref{fig : sprakelmacromolecules2}-Left, $G_0$ being the shear modulus). Note that the data point corresponding to $\sigma/G_0=3$ comes from a strain versus time measurement whose appearance is significantly different from the others (smallest applied stress of Figure \ref{fig : sprakelmacromolecules1}), which is a possible manifestation of the phenomenon of wall slipping at high stresses [\cite{deGennes2007,Shull2012}]. However, even excluding this data point, the agreement between the experimental data and the prediction is not sufficiently clear to state that the TAC model provides an explanation for the existence of the observed delayed fracture as explained in section \ref{sec : activated rupture}.

Another way to check a basic prediction coming from the  TAC-model is to measure the characteristic value $\sigma_1$ of the stress and to compare it to the expression following the model. From Eqs.(\ref{eqn : maximum}, \ref{eqn : waiting time}) $\sigma_1$ reads: 
\begin{equation}
\sigma_1=\left(\frac{6\pi}{k_BT\alpha^2} \right)^{1/4}F_s^{3/4}G_0^{1/2}.
\label{eqn : sigma_1}
\end{equation}

\begin{figure}
\begin{center}
\includegraphics[width=0.9\textwidth]{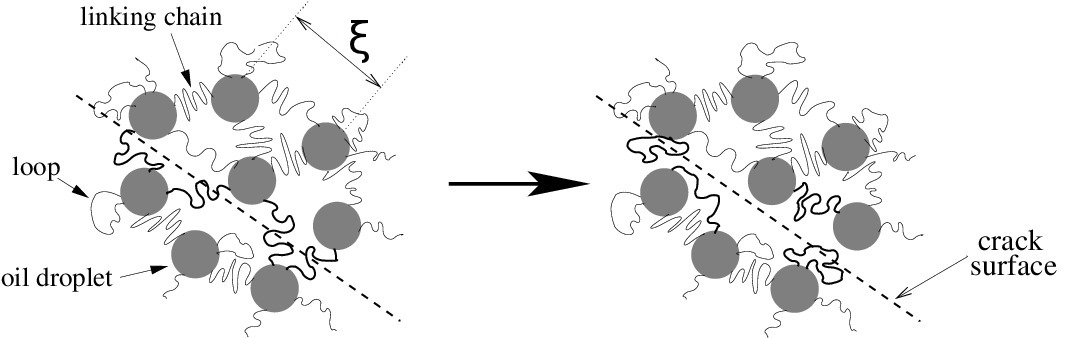}
\end{center}
\caption{Schematic of a bridged microemulsion. The telechelic polymers can either link two oil droplets or loop on a single one. (Left) Before the crack nucleation (bold dashed line) polymers can bridge oil droplets on both side of the bold dashed line. (Right) When the crack occurs, the same polymers cannot cross the bold dashed line anymore and form bridges in the other directions.} \label{fig : fracture}
\end{figure}

$\sigma_1$ can be interpreted as follows: the fracture propagation is subcritical for a load smaller than $\sigma_1$. Otherwise, the fracture nucleates instantaneously. $\sigma_1$ can be estimated as follows. Let $\xi$ be the typical mesh size of a transient network.  Following Green and Tobolsky [\cite{Green1946}], the shear modulus is $G_0 \simeq a k_BT/\xi^3$, where $a$, of order unity, depends on the network topology. In order to estimate the value of $F_s$, we consider a micro-crack of the network. First of all it is important to note that micro-cracks in these networks are completely wet by the solvent and thus only the polymer network contributes to $F_s$ (see Figure \ref{fig : fracture}). The energy cost for creating a network-solvent interface originates from the depletion of the chains monomers at the  crack interface. We use the well known analogy between transient networks and polymers solution in the semi-dilute regime: the cost in free energy per unit surface due to the depletion is [\cite{Joanny1979}] $F_s\simeq a' \frac{k_BT}{\xi^2}$, where $a'$, of order unity, also depends on the network topology. Then, Eq. \ref{eqn : sigma_1} yields:
\begin{equation}
\sigma_1\simeq  \left(\frac{6\pi a'^3a^2}{\alpha^2} \right)^{1/4}\frac{k_BT}{\xi^3}= \left(\frac{6\pi a'^3a^2}{\alpha^2} \right)^{1/4}G_0.
\label{eqn : pomeau xi}
\end{equation}
The value of the prefactor being not far from unity, it appears that $\sigma_1$ is close to the shear modulus. Particle-based simulations  where $\sigma_1$ is plotted versus polymer concentration corroborate  Eq. \ref{eqn : pomeau xi} [\cite{Sprakel2009}]. Note that in Eq. \ref{eqn : pomeau xi}, $\sigma_1$ is independent of the relaxation time and of the viscosity. It is therefore independent of the shear rate. 

The best fit  (Figure \ref{fig : sprakelmacromolecules2})  of the data of Skrezesezewska et al. [\cite{Skrzeszewska2010}] with  Eq. \ref{eqn : waiting time} (excluding, as  explained above  the data point corresponding to $\sigma = 3G_0$), gives $\frac{\sigma_1}{G_0} \sim 1.74 \pm 0.08 $, in agreement with Eq. \ref{eqn : pomeau xi}.

Tabuteau et al. have measured $\sigma_1$ using pendant drop experiments, as introduced in section \ref{sec : pendant exp}. The authors used  a model system made of oil-in-water microemulsion droplets  connected to each other by telechelic polymers (see Figure \ref{fig : fracture})[\cite{Filali1999}] at different concentrations in order to vary both the relaxation time and the elastic modulus.  
In pendant drop experiments, the increase in stress with time is large enough to neglect the probability of breaking due to a subcritical fracture  stress. Therefore the characteristic stress $\sigma_1$ in Eq. \ref{eqn : pomeau xi} is equal to the stress at the detachment in pendant drop experiments. 
The structure of this transient network being well characterized (particularly the network connectivity), it is possible to obtained an explicit expression for coefficients $a$ and $a'$ in Eq. \ref{eqn : pomeau xi} [\cite{Tabuteau2009}]  for this specific material. The agreement between measured critical stresses and the theoretical prediction is  satisfactory (Figure \ref{figure_pomeau}).

This result confirms the physical basis of the interfacial cohesive energy and  allows the authors to conclude that the observed fractures are brittle. Note that the calculated value for  the Griffith's length  is, for all the samples,  of the order of three times the mean  distance between two oil droplets. This is consistent with the Griffith's theory. Furthermore, the measured critical stress has been found to be independent of the length of the stickers and therefore of the junction energy (note that the junction energy and the interfacial cohesive energy are here independent from each other). It is therefore independent of the transient network relaxation time [\cite{Tanaka1992}]. This is consistent with the prediction of the  the TAC-model for $\sigma_1$, where neither the junction energy nor the relaxation time do appear (see Eq. \ref{eqn : pomeau xi}).

This last point is in contrast with  the experiments of Skrezesezewska et al [\cite{Skrzeszewska2010}] (see for instance Figure \ref{fig : sprakel4}), showing  that the TAC model fails -at least in part- for the experimental system used by  these authors. This led them to focus on an alternative model, based on a more microscopic description of the phenomenon of fracture nucleation: the {\em activated bond rupture model} (ABR  model).

\subsection{Activated bond rupture model (ABR model)} \label{sec : activated rupture}
Another way to describe fracture nucleation in transient networks is based on activated bond rupture [\cite{Evans1997,Chaudhury1999,Skrzeszewska2010}].

In the systems we have in mind, the polymers are not entangled,  so the single relevant time scale comes from the debonding kinetics.
The free energy variation  $\Delta {\cal G}_0$ characterizing a network junction is assumed to depend on the solvophobic length and the solvent quality. There is, by thermal fluctuations, a finite probability that a junction acquires sufficient energy to overcome the activation barrier $\Delta {\cal G}_0$ and detach spontaneously. Following the model of Green and Tobolsky [\cite{Green1946}], the exit rate ($1/\tau_0$) is estimated as the product of a natural thermal vibration frequency, $\nu$ ($\sim 10^{10}-10^{12}$ Hz) with  $\exp\left(-\Delta {\cal G}_0 /k_BT \right)$. Hence, one expects 
\begin{equation}
\frac{1}{\tau_0}=\nu \exp\left(-\frac{\Delta {\cal G}_0}{k_BT} \right).
\label{eqn : energy barrier}
\end{equation}

From the Eyring kinetic theory of fracture [\cite{Eyring1936,Kausch1978}], Tanaka and Edwards [\cite{Tanaka92_1}] postulated that the lifetime of a junction also depends on the force carried by a link, whose stored energy can supply, through mechanical work, a portion of the activation energy. The junction lifetime $\tau$ becomes:
\begin{equation}
\tau={1 \over \nu} \exp\left({(\Delta {\cal G}_0 -f\delta ) \over k_BT}\right)=\tau_0 \exp\left(-\frac{f\delta }{k_BT} \right),
\label{eqn : lifetime}
\end{equation}
where $f$ is the force carried by the link under tension, and $\delta$ a typical length on which the force works (for instance $\delta$ is the length of the solvophobic ends of telechelic polymers). 
A direct implication of this assumption is that a stretched bond breaks more quickly than the non-stretched ones. Tanaka and Edwards predict the mechanical behavior of such transient networks under different conditions (strain oscillatory, transient, constant shear rate, etc.) [\cite{Tanaka92_1,Tanaka92_2,Tanaka92_3}]; they also predict non-linear effects (such as shear thinning) observed experimentally [\cite{Tam1998,Ma2001,Sprakel2008}]. These non-linearities come from the reduction of bond lifetime under load (Eq. \ref{eqn : lifetime}). However, because this theory is itself a mean-field theory with affine deformations of the network junctions, the existence of a fracture completely escapes this model.\\

\subsubsection{Constant load} \label{sec : activated constant}

If the length of a pre-existing crack is larger than the Griffith's length, then the fracture develops instantaneously, without  delay. We are considering here, an opposite  situation, where the crack size is smaller than the Griffith's length. The crack then grows slowly and can be described by the model of  {\em the subcritical crack growth} [\cite{Santucci2007}]. Because of the presence of heterogeneities due to thermal fluctuations in the network, the stress is inhomogeneous in the material, and then the force carried by a bond is not homogeneous. It is stronger where the bond density  is low. Since the stress is concentrated near the crack tips, cross-links located just near a given  tip will experience a higher force and then a higher dissociation rate. The kinetics of breakage is accelerated, which will result in a shorter time for the size of the micro-crack to reach the Griffith length. Once the crack reaches the Griffith length a catastrophic fracture follows. 
As a first approximation (i.e. neglecting the creation of new bonds), the average time $\langle t_2 \rangle$ required for the micro-crack to reach the Griffith length has to decrease exponentially with the applied stress [\cite{Brenner1962,Zhurkov1965}]:
\begin{equation}
\langle t_2 \rangle \sim \exp\left(-\frac{\sigma}{\sigma_2} \right),
\label{eqn : sigma2}
\end{equation}
where the characteristic stress $\sigma_2$ incorporates the factor $\delta /(k_BT)$ but also takes into account the stress concentration near the preexisting crack. Note that the $\sigma^{-2}$-variation of the Griffith length (Eq. \ref{eqn : griffith length}) has been neglected because it varies much more slowly than the exponential.

Such dependence of the fracture waiting time (Eq. \ref{eqn : sigma2}) as a function of the applied stress has been observed in several types of colloidal gels [\cite{Divoux2010,Sprakel2011b}]. Skrzeszewska et al. [\cite{Skrzeszewska2010}] have checked the validity of this approach for transient networks subjected to a constant load. They found a good agreement between Eq. \ref{eqn : sigma2} and their experimental data, as shown in Figure \ref{fig :  sprakelmacromolecules2}-Right. However, it is difficult to conclude from these data what is the relevant model among  the TAC-model and the {\em activated bond rupture model} (ABR model); Indeed, none of the two plots of Figures \ref{fig :  sprakelmacromolecules2} clearly prevails over the other.

Furthermore, the range of validity of Eq. \ref{eqn : sigma2} is limited, since cross-links re-formation are not taken into account. It can therefore be so inferred that this model cannot be applied for small loads, typically when the fracture waiting time $\langle t_2 \rangle$ exceeds the characteristic time of cross-links re-formations. In addition, Eq. \ref{eqn : sigma2} also predicts a smooth variation  without any abrupt jump for the fracture waiting time regardless of the value of the applied stress, contrary to other models (see sections \ref{sec : thermally activated} and \ref{sec : spring} and Figure \ref{fig : comparaison}). This is in opposition with the existence of a experimentally observed fracture threshold (an unstressed transient network does not spontaneously break).
 
\subsubsection{Constant shear rate}

\begin{figure}[!h]
\begin{center}
\includegraphics[width=1.2\textwidth]{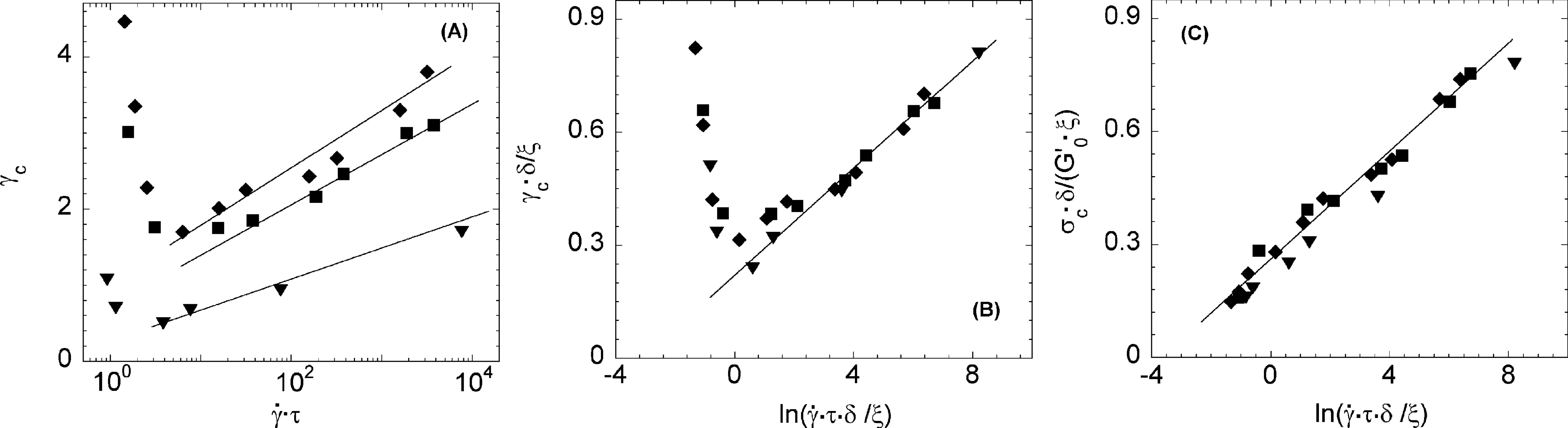} 
\end{center}
\caption{From [\cite{Skrzeszewska2010}]: A- Critical strain as a function of shear rate for transient polymer networks formed by telechelic polypeptides at different concentrations. Lines corresponds to the predictions of the ABR model. (B) Renormalized critical strain as a function of the logarithm of a renormalized shear rate, (C) Same for critical stress.}
\label{fig : sprakel4}
\end{figure}

Sections \ref{sec : thermally activated}  and \ref{sec : activated constant}  deal with crack nucleation for transient networks subjected to constant loads. In this case,  the sample flows before breaking because the fracture does not appear instantaneously and because the material is a fluid. In this situation, the shear rate  varies with time because it is subject to fluctuations even when the steady flow is still established. \\
Very commonly, a shear rate is imposed in rheological tests. Some studies have focused on fracture nucleation  in a transient network subjected to a constant shear rate (see e.g. section \ref{sec : stationary}) [\cite{Berret2001,Skrzeszewska2010}]. Skrzeszewska et al. [\cite{Skrzeszewska2010}] observed that the failure stress and the waiting time (before the fracture occurs) depend on the imposed shear rate (Figure \ref{fig : sprakel4}): the waiting time is found to decrease and the critical stress to increase as the constant shear rate is higher. Interestingly, the critical strain exhibits a non monotonic behavior  on the share rate (Figure \ref{fig : sprakel4}a), very similar as what is observed  in uniaxial extension of polymeric liquids [\cite{Malkin1997}]. Particle-based simulations also show a weak but significant increase of the fracture stress with applied overall shear rate [\cite{Sprakel2009}]. The authors of [\cite{Skrzeszewska2010}] have used the ABR model to explain these results. Within a mean field approximation, the deformation is affine and a bridging chain in the network is stretched at a speed $\xi \dot{\gamma}$, where $\dot{\gamma}$ is the (macroscopic) applied shear rate and $\xi$ the typical distance between junctions. Supposing that bridging chains act as Gaussian chains with a spring constant $k_BT/\xi^2$, the force carried by a cross-link increases as $f(t)=K\frac{k_BT}{\xi}\dot{\gamma}t$, where the factor $K$ accounts for the local stress enhancement in the vicinity of the crack tip [\cite{Skrzeszewska2010}]. With Eq. \ref{eqn : lifetime} one obtains, for the survival probability of a bond $P(t)$ [\cite{Evans1997}]:
\begin{equation}
P(t)=\exp\left\{-\int_0^{t}\mbox{d} t' \frac{1}{\tau_0} \exp\left(\frac{K k_BT\dot{\gamma}t'\delta }{\xi k_BT} \right)\right\}.
\label{eqn : sprakel approx}
\end{equation}
The average lifetime of a cross-link is
\begin{equation}
\langle \tau_0' \rangle = \int_0^{\infty}P(t)\mbox{d}t.
\label{eqn : proba}
\end{equation}
 In the high shear-rate limit, $\dot{\gamma}\tau \gg \frac{\xi}{K\delta}$ and $\dot{\gamma} \tau \gg 1$,  re-formation of cross-links can be neglected. Eq. \ref{eqn : proba} simplifies and it follows that a fracture occurs after a time $t_2' \simeq \langle \tau_0' \rangle$
\begin{equation}
 t_2'  = \frac{\xi}{K\delta \dot{\gamma}}\ln\left(\frac{K\delta  \dot{\gamma}\tau}{\xi}\right),
\end{equation}
and the critical strain reads $\gamma'_2=\dot{\gamma}t_2'\simeq  \frac{\xi}{K\delta}\ln\left(\frac{K\delta \dot{\gamma} \tau}{\xi}\right)$. Neglecting a possible strain hardening [\cite{Tabuteau2009b,Serero2000}], the stress at the rupture ($\sigma_2'$) can be estimated as $\sigma \simeq G_0 \gamma$. The experimental critical strain $\gamma_2'$ is plotted in Figure \ref{fig : sprakel4}-A for three different concentrations of a transient polymer networks formed by telechelic polypeptides (from [\cite{Skrzeszewska2010}]).  In order to make a more quantitative comparison between their experiments and the model, the authors  have plotted dimensionless values of $\gamma_2'$ and $\sigma_2'$  ($\gamma'_2\delta/\xi^3$ and $\sigma_2' \delta/(G_0\xi)$) as a function of $\ln(\dot{\gamma}\tau \delta/\xi)$, where $\delta$ is the estimated length of a junction and $\xi$ is estimated using $G_0 \simeq k_BT/\xi^3$. The curves superimpose with a nice agreement with theoretical  predictions of Eq. \ref{eqn : proba} in the high shear limit. This gives  a strong argument in favor of the relevance of the ABR model for this system, whereas the TAC-model predicts a shear rate independent failure stress.  However, these predictions does not seem applicable to other transient networks, where a shear stress independent failure stress has been reported [\cite{Tabuteau2009}]. As explained below, the main weakness of the ABR model is that cross-links re-formations are neglected. 

\subsection{Self healing and activated bond rupture nucleation (SH-ABR model)} \label{sec : spring}

The fracture waiting time predicted by the ABR model leads to a finite waiting time whatever is the applied stress, even if it is zero. Following this prediction, an unstressed transient network should spontaneously break after a finite time. This inconsistency arises because the creation of new bonds (which could replace the broken bonds) is not taken into account. Recently, Mora [\cite{Mora2011}] has proposed a model, the  {\em Self healing and activated bond rupture nucleation model} (SH-ABR model), which aims to describe fracture nucleation in transient networks considering {\bf both} activated bond rupture and cross-links re-formation to overcome the intrinsic limitation of the ABR model.

\begin{figure}[h]
\begin{center}
\includegraphics[width=0.5\textwidth]{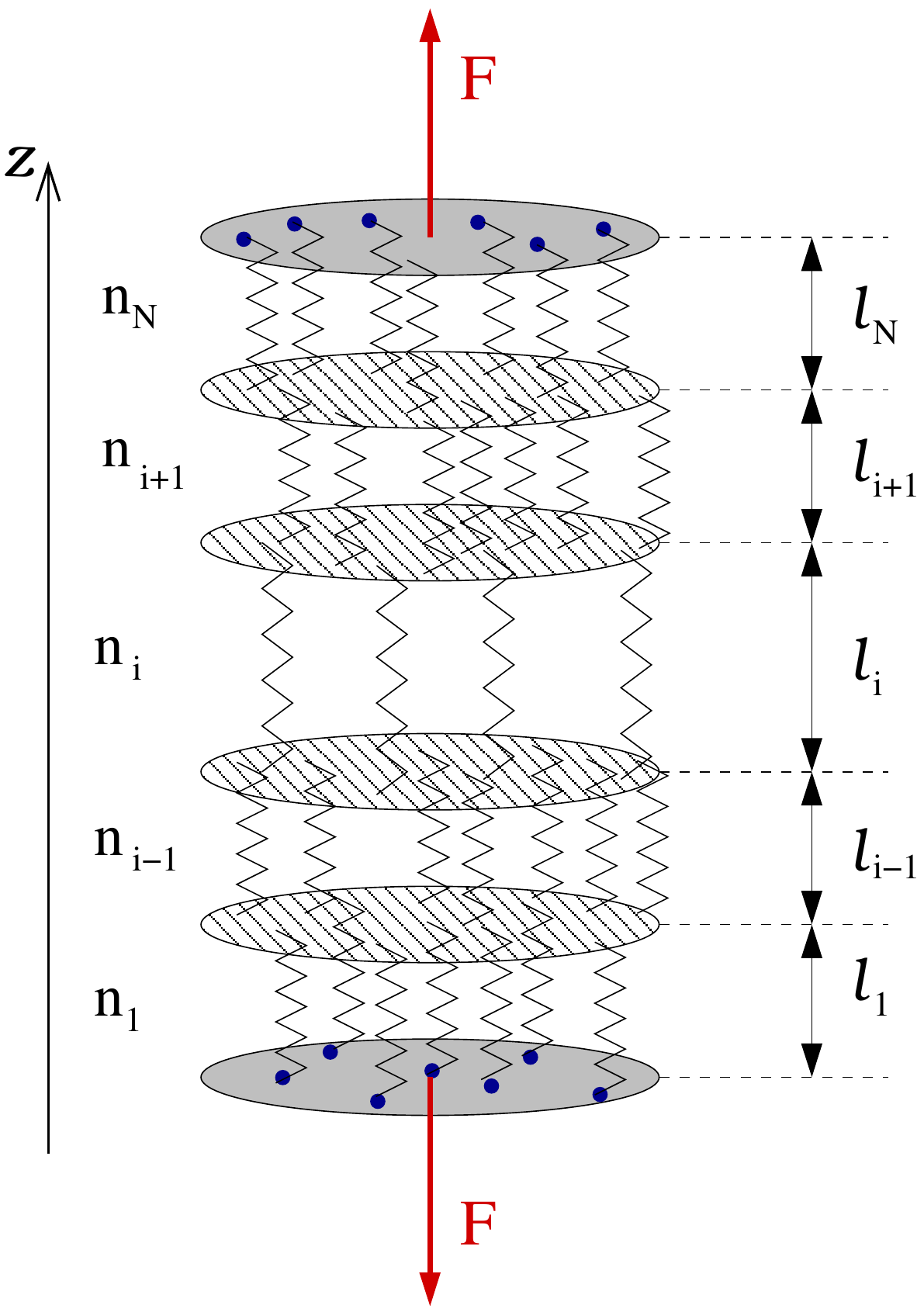}
\end{center}
\caption{From [\cite{Mora2011}]: Schematic of the one-dimensional model. The system is composed of $N$ layers. The layer number $i$ is connected to the layer number $i +1$ by  $n_i$ bonds. The total number of bonds and layers is fixed once and for all. All bonds are identical. They can connect only two adjacent layers. One end of a bond can migrate from one layer to another. A force applied to the ends is transmitted layer by layer. Within a layer, this force is split evenly between the springs.
 } \label{fig : system 1d}
\end{figure}

It is based  on the fiber bundle model, usually used to study fracture in heterogeneous materials [\cite{Coleman58,Moreno99,Kun2000,Guarino2001,Pradhan2010}]: a set of fibers (or bonds) is located on a supporting lattice and a random stress threshold is  assigned to its elements. The set is loaded and a bond breaks when its load exceeds a threshold value. In the equal load sharing hypothesis (or global), which is the simplest scheme one can suppose, it is assumed that the load carried by failed elements is equally distributed among the surviving elements of the system. The SH-ABR model deals with systems without entanglement. The relaxation time of a link is then assumed to be far shorter than the bonds lifetime.In this model, a bond just once broken becomes active again: It can either take back the place it had or bridge another junction. Mora [\cite{Mora2011}] first considered a one dimensional model (see Figure \ref{fig : system 1d})  subjected at time $t=0$  to an external stress $\sigma$. It is assumed that (i) following the Eyring kinetic theory of fracture [\cite{Eyring1936,Kausch1978,Tanaka92_0}], the lifetime $\tau$ is given by Eq. \ref{eqn : lifetime}, and
(ii) one end of a bond that has just escaped will reconnect over a much shorter duration than the (load depending) lifetime of a bond (as in most of experimental systems).

An analytical  critical magnitude of the applied stress is predicted in 1D: 
\begin{equation}
\sigma_3=\frac{n_0 \xi k_BT}{\delta },
\label{eqn : force critique 1d}  
\end{equation}
where $n_0$ is the number of network junctions per unit volume and $\xi$ the mesh size of the network (the distance between two layers in the one-dimensional model, see Figure \ref{fig : system 1d}). Beyond this threshold, any concentration inhomogeneity  in the bonds distribution is amplified, leading to the formation and the catastrophic and irreversible growth of cracks. On the contrary, any inhomogeneity in the bonds distribution decreases in time when the applied stress is below this threshold value. Note that the load dependence of the bond lifetime  (formally: $\delta \neq 0$ in Eq. \ref{eqn : force critique 1d}) is here a necessary condition for the existence of a fracture stress threshold.

\begin{figure}[!h]
\begin{center}
\includegraphics[width=0.7\textwidth]{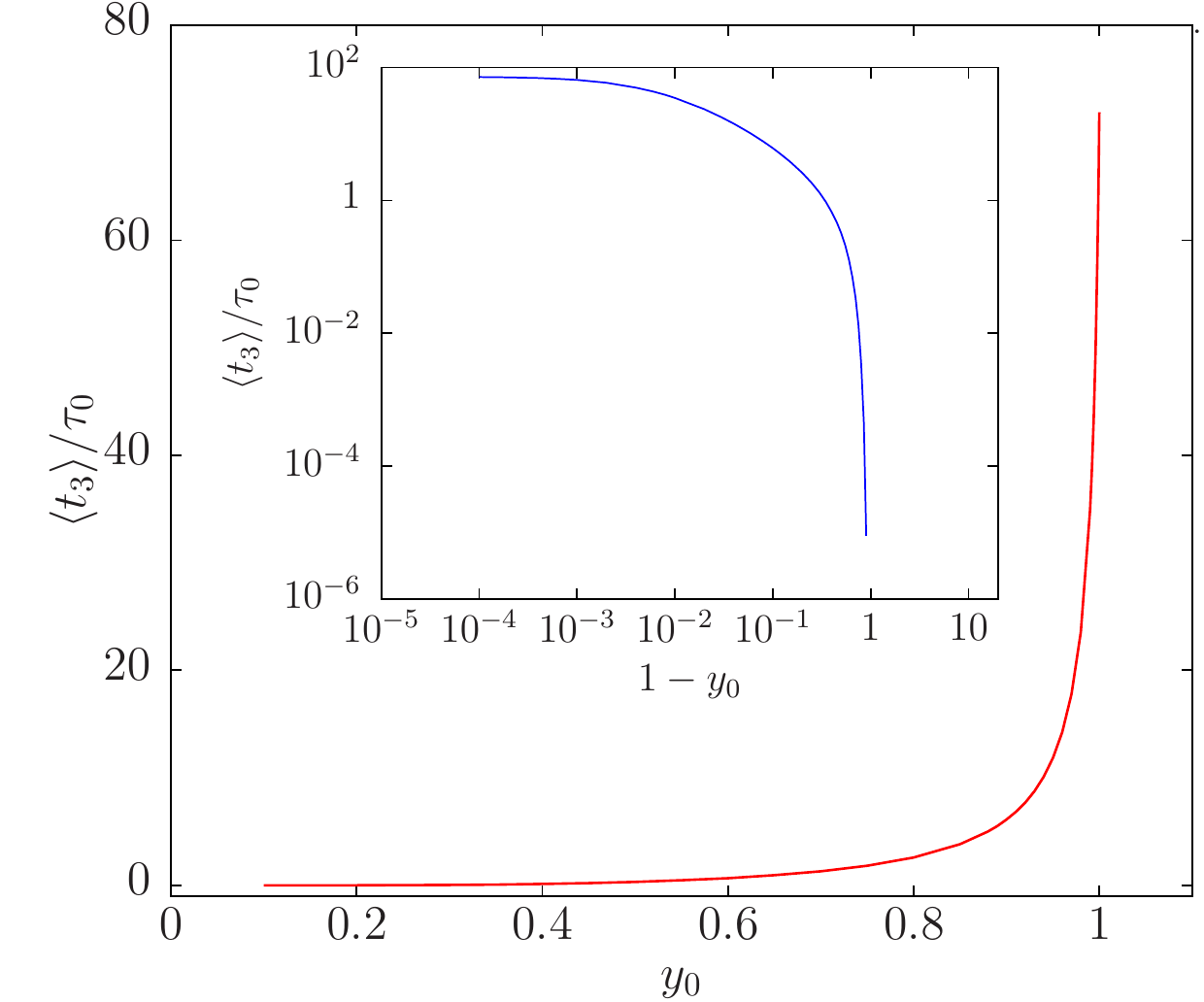} 
\end{center}
\caption{From [\cite{Mora2011}]: Reduced failure time $t_{3}/\tau_0$ as a function of $y_0=\frac{n_0\xi k_BT}{\sigma \delta}$. This time diverges to infinity on approaching the rupture threshold. Insert: Same curve in logarithmic scales.}
\label{fig : time_vs_stress}
\end{figure}

The time $\langle t_3\rangle$ at which the concentration of bonds vanishes somewhere in the one-dimensional system is by definition the failure time. This time is found to depend on the ratio $\frac{n_0\xi k_BT}{\sigma \delta}$ (Figure \ref{fig : time_vs_stress}). It tends to infinity as $\frac{n_0\xi k_BT}{\sigma \delta }$ tends to one {\em i.e.} when approaching the failure threshold ($\sigma=\sigma_3$). Of course, the failure time is not defined for $\frac{n_0\xi k_BT}{\sigma \delta}> 1$.

A similar result was found when the strain is imposed. The threshold value of the deformation beyond which rupture occurs is simply equal to the ratio of the critical stress (Eq. \ref{eqn : force critique 1d}) with the elastic modulus of the system. The fracture waiting time is finite when the imposed strain is above this threshold. It decreases gradually as the imposed strain increases.

Numerical simulations are used to validate the results, at least semi-quantitatively, for two-dimensional networks [\cite{Mora2011}]. Note that the fractures under a constant shear rate have so far not been studied within this model.

In conclusion, the introduction of a self-healing process in the ABR model leads to a true fracture stress threshold. 

\subsection{Ranges of validity and discussion} \label{sec : validity}

\begin{figure}[!h]
\begin{center}
\includegraphics[width=0.8\textwidth]{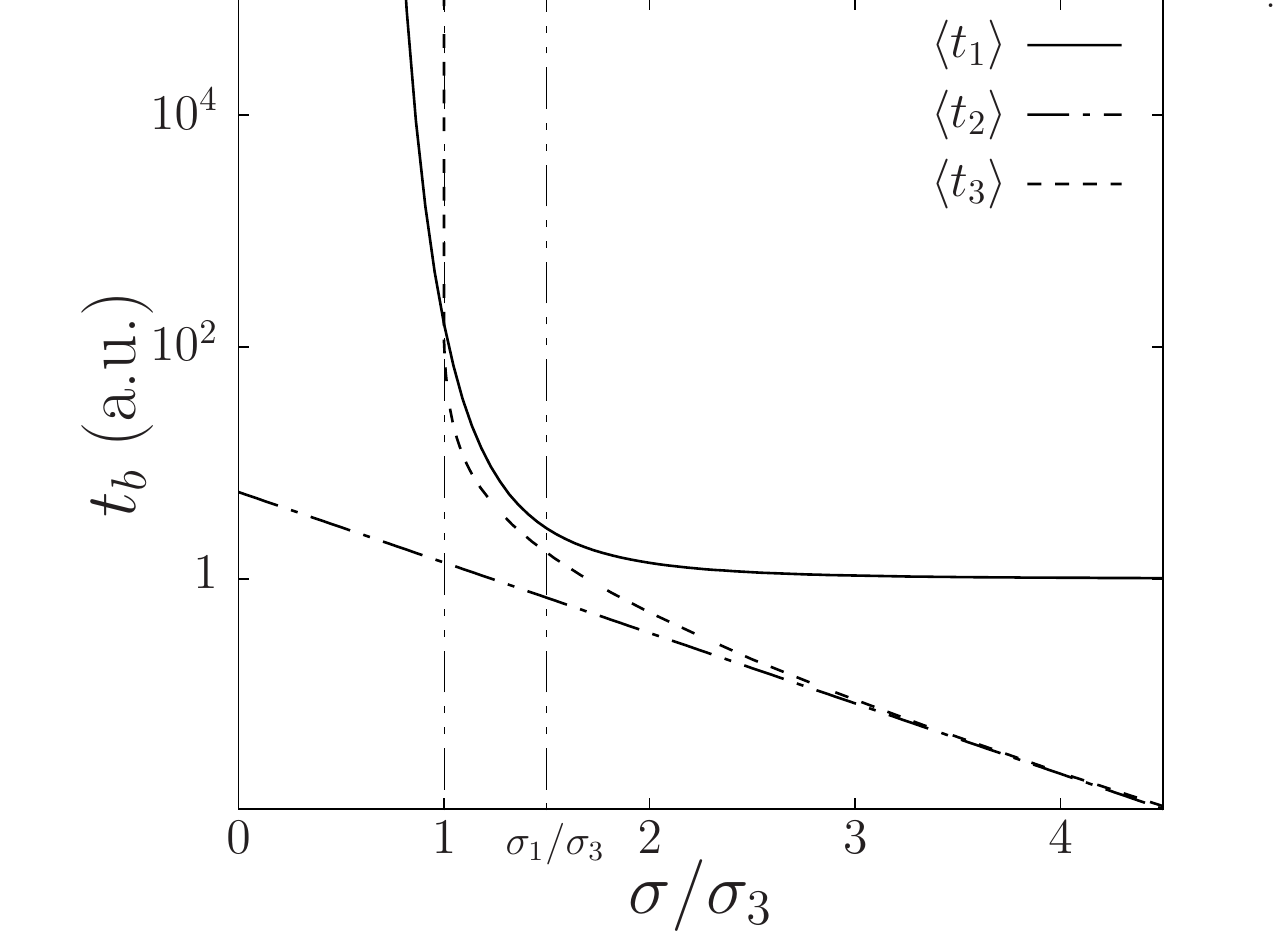}

\includegraphics[width=0.8\textwidth]{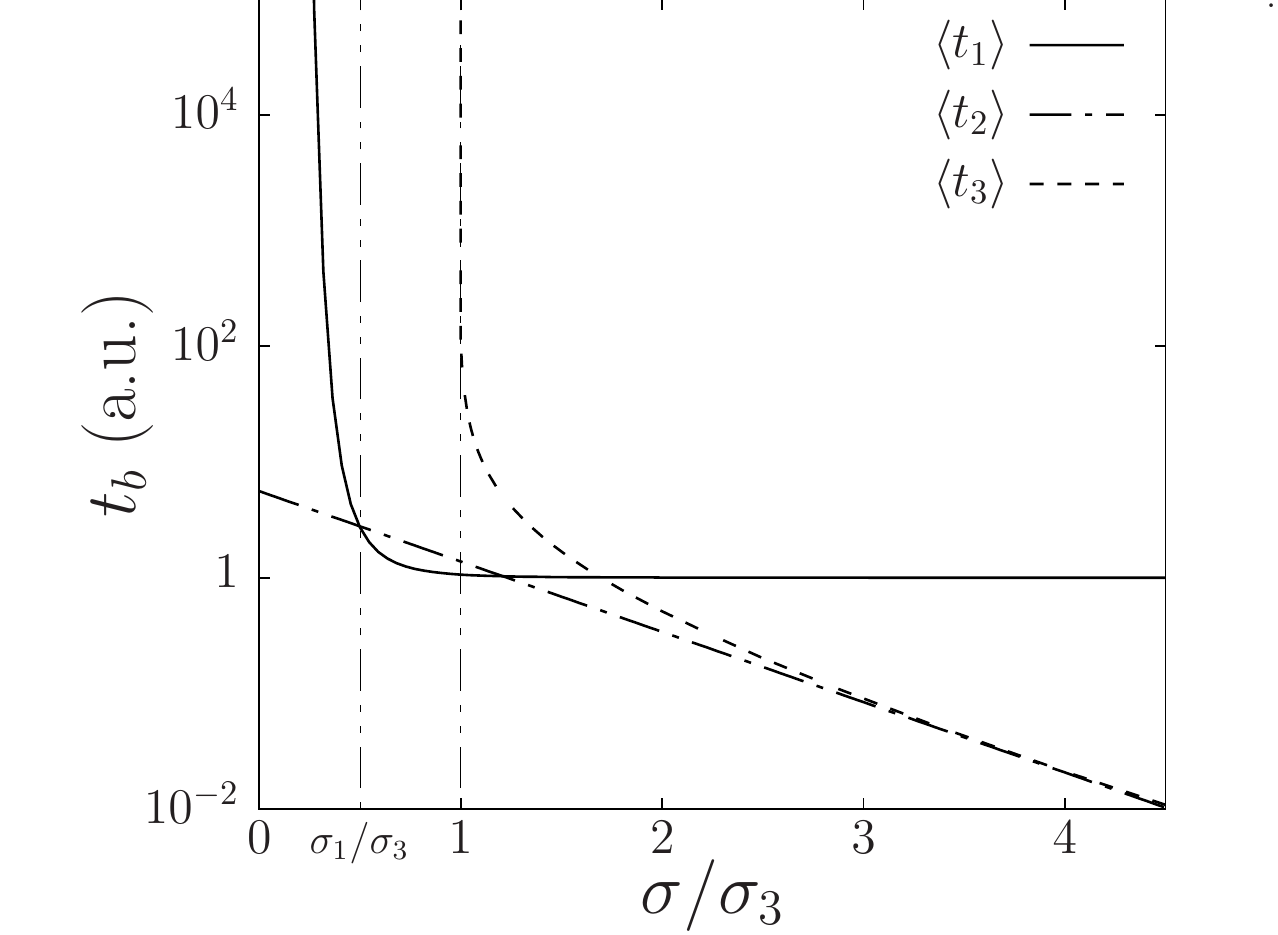}  
\end{center}
\caption{Fracture waiting time $t_{b}$ as a function of the applied stress calculated for the three models. Top: $\sigma_1>\sigma_3$. In this case, the curves giving  $\langle t_1\rangle$ and $\langle t_3\rangle$ intersect and the corresponding waiting time is not molecular ($t_b \gg \tau_1$). Bottom: $\sigma_3>\sigma_1$. $\langle t_1\rangle$ and $\langle t_3 \rangle$ intersect at a molecular time ($\tau_1$ in Eq. \ref{eqn : waiting time}). }
\label{fig : comparaison}
\end{figure}

We have presented three models that lead to different predictions for the fracture waiting time and for the critical failure stress (see Figure \ref{fig : comparaison}).

When the applied stress is large, the average bond lifetime is very short (see Eq. \ref{eqn : lifetime}) and the effects of cross-links re-formation become negligible. Here, {\em large} means $\exp(-f\delta/(k_BT)) \ll 1$, i.e. $\sigma \gg \sigma_3$ (where $\sigma_3$ is given by Eq. \ref{eqn : force critique 1d}). Figure \ref{fig : comparaison} shows that for $\sigma \gg \sigma_3$  the decay of the fracture waiting time as a function of the stress is exponential, in accordance with the prediction of the ABR model, i.e. $\langle t_3\rangle  \simeq \langle t_2 \rangle$.  So in this limit, the ABR and the SH-ABR models lead to the same predictions.

In the opposite limit, where  $\exp(-f\delta/(k_BT)) \lesssim 1$ or equivalently $\sigma \lesssim \sigma_3$, the effects coming from the load dependence of the cross-links lifetime are negligible (formally, $\delta = 0$ in Eq. \ref{eqn : lifetime}). The ABR model is no longer valid because the cross-links re-formation events are important. In this case the SH-ABR model predicts the absence of fracture (because $\sigma < \sigma_3$). Within this limit, the TAC-model is the only one able to explain the origin of the experimentally observed fractures.

Because $\langle t_3 \rangle \rightarrow \infty$ (resp. $0$) for $\sigma \rightarrow \sigma_3^-$ (resp. $\infty$) and $\langle t_1 \rangle \rightarrow \infty$ (resp. $\tau_1$) for $\sigma \rightarrow 0$ (resp. $\infty$), $\langle t_1 \rangle $ is smaller  (resp. larger) than $\langle t_3 \rangle $ for low (resp. high) stresses (Figure \ref{fig : comparaison}).

Let us suppose that the relevant model is the one which predicts the shortest fracture waiting time for a given value of the stress. 
If $\sigma_1<\sigma_3$,  then $\langle t_1 \rangle$ intersects $\langle t_3 \rangle $ for a stress larger than $\sigma_1$ (Figure \ref{fig : comparaison}-right). As a result,  $\langle t_3 \rangle $ will be the relevant fracture waiting time for stresses for which $\langle t_1 \rangle $ is of order $\tau_1$ (a molecular time, Eq. \ref{eqn : waiting time}), in contradiction with the implicit hypothesis of a finite breaking time. In the opposite case, where $\sigma_1>\sigma_3$ (Figure \ref{fig : comparaison}-left), $\langle t_1 \rangle$ intersects $\langle t_3 \rangle $ for a  stress smaller than $\sigma_1$ for which $\langle t_1 \rangle$ is a priori far larger than $\tau_1$. The SH-ABR model then predicts a macroscopic fracture waiting time smaller than $\langle t_1 \rangle$. It is so the relevant model in this case.

  The  TAC and SH-ABR models described above must be thought of as asymptotic models. For intermediate stresses a more complete description must take into account all the ingredients of both models to give a more refined  transition between the  TAC-model and the SH-ABR model.

  The condition for which the SH-ABR model is likely to adequately describe the fracture waiting time is (i) $\sigma_1>\sigma_3$ and (ii) the applied stress is greater than the crossover stress for $\langle t_1 \rangle$ and $\langle t_3 \rangle$. From Eq. \ref{eqn : pomeau xi} and \ref{eqn : force critique 1d} condition (i) yields:
\begin{equation}
 \left(\frac{6\pi a'^3a^2}{\alpha^2} \right)^{1/4} G_0>\frac{n_0 \xi k_BT}{\delta}\simeq \frac{\xi}{\delta}G_0.
\label{eqn : comparaison total}
\end{equation}
Eq. \ref{eqn : comparaison total} leads to the approximate condition (omitting the prefactor of order unity):
\begin{equation}
\delta > \xi.
\label{eqn : comparaison simple}
\end{equation}



The TAC-model applies at any stress when $\sigma_3>\sigma_1$, i.e. for $\xi \gg \delta$. \\

Let us come back to the two main experimental systems introduced above. In the experiments of Tabuteau et al. [\cite{Tabuteau2008,Tabuteau2009}], the mean distance between junctions is of the order of $\xi\sim 130 ~nm$ whereas the length of a hydrophobic sticker is about $\delta \sim 10 ~mm$. Thus, $\delta/\xi \ll 1$ is consistent with the experimental measurements in agreement with the TAC-model. 

Because the fine structure of the triple-helical junction in transient polymer networks formed by telechelic polypeptides as studied by Skrzeszewska is not well known, the estimation of $\delta/\xi$ is less accurate. The authors took $\delta \simeq 9~nm$, leading to  a ratio $\delta/\xi$ ranging from 0.18 to 0.5, depending of protein concentration. This limit corresponds to the situation  where it is expected that  the different models should be coupled. However, the experimental results obtained by the authors lead in favor of the ABR  (or more  SH-ABR) model.  This suggests that the  ABR model is preponderant even for values  $\delta/\xi$ slightly below 1. \\
  
In summary, we have introduced three models, each of them predicting a fracture waiting time as a function of the applied load. We have assumed that the relevant model is the model predicting the shortest waiting time. Then, it is possible to predict the variation of the waiting time. The threshold values for the applicability of these models depend on (i) the value of the applied stress, and on (ii) the ratio $\delta/\xi$.

    
    \section{Conclusion and Perspectives}
Transient networks appear as simple model systems for the physicist to achieve a fundamental understanding of fracture phenomenons in complex fluids because of the unique transient character of the polymeric bonds; In this review we have focused on the  specificities arising from the  fluid and self-healing characters of these materials. 
Fractures of transient networks have been reported in several configurations: shear, extensional or hybrid geometries. The main results include:
 
 $\bullet $ Experimental  evidences of the occurrence of  true fractures, that can be unambiguously distinguished from hydrodynamic instabilities in various geometries.

 $\bullet $ Measurements of failure thresholds, expressed as the applied stress or the applied deformation (e.g. shear rate), and  measurements of the fracture waiting time for subcritical stresses or deformations.
Fracture waiting time as a function of the applied stress and threshold fracture stresses have been measured by different groups. These measurements have led to different conclusions in apparent contradiction.  Different and somewhat contradictory  theoretical approaches developed to understand the experiments have been reviewed. Two are emerging among them: the \emph{Thermally activated crack nucleation model} (TAC model) and the  \emph{Self healing and activated bond rupture nucleation model } (SH-ABR model). Each model has a limited range of validity. Two characteristic lengths play a key role to discriminate between them: the average distance between two junctions of the network ($\xi$), and the size of the adhering part of the junctions ($\delta$). For $\xi \gg \delta$, the TAC model is always relevant, whereas for $\xi \ll \delta$,  the  TAC-model is relevant for low applied stresses and the SH-ABR model is relevant for high stresses. This  should explain the diversity of the experimental behaviors observed for different experimental systems. \\
An interesting challenge will be to formulate experimental systems for which the ratio $\delta/\xi$ can be finely tuned to check this hypothesis.

 $\bullet $ Most of the  experimental results and theoretical analysis have dealt  situations where the lifetime of bonds is large  with respect to the relaxation time of the polymeric linkers and so, dominates the systems response. This  corresponds to brittle  fractures: the  understanding of the physics involved in nucleation  and of brittle fractures begin to be clarified. The opposite situation where the relaxation of the polymeric linkers  is comparable or larger than lifetime of the bonds is still poorly understood. Transient networks  with polymer links in the entangled regime, or with nodes of tunable morphology, or near the percolation transition (marginal transient networks) will  provide experimental realizations of this situation. From the theoretical side, this new time scale  could be included in the SH-ABR model. This future direction will tackle dissipation mechanisms (viscous dissipation, structural reorganization, etc.) involved during nucleation of ductile fractures and during fracture propagation, to progress  in the fundamental understanding of the concepts of brittleness and ductility in complex fluid.
 
 $\bullet $ Finally, the role of the rigidity of the  polymeric linkers in the fracture of transient networks is a big and challenging question to address in the future.

\begin{acknowledgements}
This work has been supported by ANR under Contract No. ANR- 2010-BLAN-0402-1 (F2F)
\end{acknowledgements}

\bibliographystyle{spbasic}      
%
%

\end{document}